\documentclass[rmp,twocolumn,showpacs,floatfix,amsmath,amssymb]{revtex4-1}
\usepackage{graphicx}
\usepackage{amsmath,amssymb,amsfonts}
\usepackage{color,bm}
\usepackage[T1]{fontenc}
\usepackage[colorlinks=true ,citecolor=blue]{hyperref}

\def\fref#1{figure~\ref{#1}}
\def\Fref#1{Figure~\ref{#1}}
\def\beq{\begin{equation}}
\def\eeq{\end{equation}}
\def\beqn{\begin{eqnarray}}
\def\eeqn{\end{eqnarray}}

\renewcommand{\bf}{\mathbf}

\def\Tr{\mathrm{Tr}}
\def\rme{\text{e}}
\def\rmi{\text{i}}
\def\bi#1{\bm{#1}}

\begin{document}
\title{Quantum transport in Rashba spin-orbit materials: A review}

\author{Dario Bercioux}
\email{dario.bercioux@dipc.org}
\affiliation{Donostia International Physics Center (DIPC), Manuel de Lardizbal 4, E-20018 San Sebasti\'an, Spain}
\affiliation{IKERBASQUE, Basque Foundation of Science, 48011 Bilbao, Basque Country, Spain}
\affiliation{Dahlem Center for Complex Quantum Systems and Institut f\"ur Theoretische Physik, Freie Universit\"at Berlin, Arnimallee 14, D-14195 Berlin, Germany}

\author{Procolo Lucignano}
\email{procolo.lucignano@spin.cnr.it}
\affiliation{CNR-SPIN, Monte S.Angelo -- via Cinthia,  I-80126 Napoli, Italy}
\affiliation{Dipartimento di Fisica, Universit\`a di Napoli ``Federico II'', Monte S.Angelo, I-80126 Napoli, Italy}

\begin{abstract}
In this review article we describe spin-dependent transport in materials with spin-orbit interaction of Rashba type. We  mainly focus on semiconductor heterostructures, however we consider topological insulators, graphene and hybrid structures involving superconductors as well.  We start from the Rashba Hamiltonian in a two dimensional electron gas and then describe transport properties of two- and quasi-one-dimensional systems. The problem of spin current generation and interference effects in mesoscopic devices is described in detail. We address also the role of Rashba interaction on localisation effects in lattices with nontrivial topology, as well as on the Ahronov-Casher effect in ring structures. A brief section, in the end,  describes  also  some related topics including the spin-Hall effect, the transition from weak localisation to weak anti localisation and the physics of Majorana Fermions in hybrid heterostructures involving Rashba materials in the presence of superconductivity.   
\end{abstract}
\maketitle

\tableofcontents


\section{Introduction}

All the modern electronic devices are based on a classical paradigm of negative carriers ---  electrons --- and positive charge carriers ---  holes. This leaves completely unused an additional degree-of-freedom carried by both carriers: the spin.
Spintronics explores phenomena that interlink the spin and the charge degree-of-freedom. It is the field where traditional solid-state physics and material research have created their strongest bond, with each taking alternate leading roles.
From our perspective, spintronics can be mainly divided into two distinct subfields: metal and  semiconductor spintronics~\cite{zutic:2004,Sinova:2012}.

Metallic spintronic devices originate from the discovery of giant magnetoresistance (GMR) in 1988~\cite{Baibich:1988,Binasch:1989} and the subsequent development of the spin valve~\cite{Dieny:1991}. The GMR effect can be understood by assuming that any spin current  is carried by two ``types" of carriers, spin-up and spin-down. The two-channel picture of spin transport proposed by Mott explains the behaviour of magnetoresistive devices~\cite{Camley:1989}, including GMR and tunnelling magnetoresistance (TMR)~\cite{Moodera:1995}, as well as spin injection into metals~\cite{johnson:1988}.

In the GMR effect, two ferromagnetic layers sandwich a non-ferromagnetic metal layer  of nanometer thickness. When the magnetisation of the two ferromagnetic layers is parallel, the valve is in a low resistance state (open). When the two are antiparallel, the valve is in a high resistance state (closed). The TMR effect rather than the GMR one, takes place when a metallic spacer is substituted by an insulating barrier. In  TMR at room temperature, the spin-vale effect increases  by a factor of ten with respect to  GMR. This is the basic principle of hard disk read heads and magnetic random access memories. These are some of the most successful technologies of the past decades, with scaling trends outdoing even complementary metal-oxide-semiconductor, the  technology for constructing integrated circuits. Albert Fert and Peter Gr\"unberg have been awarded  the Physics Nobel Price in 2007 for their studies on the GMR effect.

Semiconductor spintronics, on the other hand, has the potential to be integrated seamlessly with nowadays' semiconductor electronics. It is progressing along a similar path  as  
metallic spintronics and has achieved a remarkable success in the past decade. The spin-field effect transistor (spin-FET) was proposed in 1990~\cite{datta:1990} as a development to the GMR set-up where the two ferromagnets are left collinear and the ``on" and ``off" status is achieved by rotating the injected spin when travelling through the two contacts. 
This geometry does not require a magnetic field for switching the relative magnetisation of the two ferromagnets, thus reducing spurious magnetic fields into the electronic circuits. The spin-FET is an electronic analog of an electro-optical modulator for photons. A ferromagnet injects spin polarised carriers into a semiconductor channel. Here, due to the modulation of the spin-precession length, it is possible to vary the spin orientation. This modulated spin signal is then detected by a second ferromagnet. The spin precession length is varied via top- and back-gate voltages that  tune the strength of the spin-orbit interaction (SOI) of Rashba (R) type ~\cite{rashba:1960,bychkov:1984}. The RSOI originates from the lack of inversion symmetry along the growth direction of the semiconductor heterostructure that is hosting a two-dimensional electron gas (2DEG)~\cite{silva:1994,silva:1997}. It can be tuned by changing the shape of the confining potential via the application of an external electric field~\cite{nitta:1997,schapers:1998,grundler:2000,miller:2003,Meier:2007}. However, the  conductivity mismatch problem~\cite{johnson:1988,schmidt:2000} between a ferromagnet and a low-dimensional semiconductor has  hindered, for a long time,  an efficient injection of spin polarised currents. Progress for solving this problem came  with the research on ferromagnetic semiconductors~\cite{ohno:1998}. Nevertheless, only recently it has been possible to  implement completely the functionality of the spin-FET {first} in Ref.~\cite{koo:2009}, and after, with minor changes, in Ref.~\cite{Chuang:2014}.

Semiconductor spintronics is also relevant because it has permitted to observe other coherent phenomena, such as the Aharonov-Casher (AC) effect~\cite{aharonov:1984} in ring-type and other loop structures~\cite{Yau:2002} so as proposed by many groups~\cite{nitta:1999,Frustaglia:2004,Bercioux:2005b,Capozza:2005,Lucignano:2007}. The Aharonov-Casher effect is the analog of the Aharonov-Bohm (AB) effect~\cite{aharonov:1959} for particles with an angular momentum in an external electrical field~\cite{richter:2012}. In the case of carriers in a semiconductor 2DEG, the RSOI permits the coupling of the electron spin with the external electric field. Other phenomena, \emph{e.g.}, the spin-Hall effect~\cite{dyakonov:1971,Hirsch:1999,Murakami:2003,Sinova:2004}, have been predicted and observed in semiconductors~\cite{Kato:2004,Sih:2005,Wunderlich:2005a,Wunderlich:2005b,Sih:2006} before their observation in metals~\cite{Valenzuela:2006}. A key role in the success of semiconductor spintronics is the tunability of the RSOI via external gating.

The material in this review article is presented in the following form: in Sec.~\ref{materials} we present a summary of the materials with sizable RSOI effects, these range from the standard semiconductor heterostructures,  carbon-based materials,  topological insulators (TIs) to Weyl semimetals (WSs). In Sec.~\ref{transport}, in order to introduce some basic transport property of carriers in systems with RSOI, we revise the spectral properties of 2DEGs and graphene with RSOI, in addition we spend some effort to underline the basic difference in the presence of a further lateral confinement, thus moving from the  {two-dimensional (2D)}  to the quasi-{one-dimensional}-case (quasi-{1D)}. After, we introduce the concept of spin-double refraction that is a phenomenon taking place in a hybrid structure composed of region with RSOI sandwiched between two regions without RSOI. In Sec.~\ref{spin:current} we give a definition of pure spin currents and {propose a} derivation of a formula for a spin current in the standard Landauer-B\"uttiker approach, furthermore we introduce two proposals for {generating}  pure spin currents --- based on the ratchet and the pumping effect. Next, we give an in-depth analysis of the spin-FET by Datta and Das  and an explanation of its working principle, finally we present  a recent experiment showing  a possible  realisation of such a transistor. In Sec.~\ref{berry} we give a  complementary look at the effects induced by RSOI, specifically we will analyse its effects as non-Abelian gauge field. Particularly, we look at non-Abelian effects in quantum networks and quantum rings. In  Sec.~\ref{related} we give a very brief and general overview over other aspects of the physics associated to the RSOI, specifically we look at the  spin-Hall effect (SHE) in Sec.~\ref{SHE}, than, in Sec.~\ref{WAL},  we will see how RSOI can modify the weak localisation (WL) into weak anti-localisation (WAL)  in a disordered system. Finally in Sec.~\ref{MFsec} we will quickly  review the vibrant field of the quest of Majorana quasi particles in quantum wires with RSOI, magnetic field and superconductivity. In Sec.~\ref{conclusions} we present our outlook for this growing field of condensed matter physics.
\section{Materials with Rashba Spin-orbit interaction}\label{materials}

On a very general basis, SOI routes down to the relativistic correction to hydrogen-like atoms~\cite{bransden:2003,Sakurai:2014}. It reads 
%
%
\begin{eqnarray}\label{soi:atom}
\mathcal{H}_\mathrm{SO} & = & - \frac{2\hbar}{(2mc)^2} \bm{s}\cdot\left( \bm{E}(\bm{r}) \times \bm{p}\right) \\
&=& \frac{1}{r} \frac{d V(r)}{d r} \bm{s} \cdot (\bm{r}\times \bm{p}) \nonumber
\end{eqnarray}
%
%
where $\bm{p}$ is the momentum operator and $\bm{s}=\left(s_x,s_y,s_z \right)$ is a vector of Pauli matrices describing the spin $\bm{S} = \frac{\hbar}{2} \bm{s}$ operator. 
By considering a Coulomb electrostatic potential $V(r)=-\frac{Ze^2}{r}$ we have that the strength of this relativistic correction  goes as $\frac{Ze^2}{r^3}$, thus proportional to the atomic number $Z$~\cite{bransden:2003,Sakurai:2014}. From the relativistic correction expressed in Hamiltonian \eqref{soi:atom} we learn that materials characterised by a higher atomic number will present in general a stronger SOI, this will be the case for TIs and Weyl- and Dirac-semimetals. 

\subsection{Semiconductors heterostructures}
In this review we shall mostly discuss the physics of  low-dimensional semiconductor heterostructures. The simplest example is a 2DEG. 
This is commonly realised in III-V semiconductor heterostructures by modulating the doping density along the growth direction --- usually~$(001)$~\cite{Bastard:1988,Seeger:2004}.  In solids, SOI can have different nature depending on the crystal  symmetries  \cite{samokhin:2009}, however in this review article we shall mostly deal with the two  main independent sources of SOI in III-V semiconductors, namely the RSOI, due to the  lack of structural inversion symmetry (SIA)~\cite{silva:1994,silva:1997} and  a Dresselhaus (D) term due to a lack of bulk inversion symmetry (BIA)~\cite{dresselhause:1955}. 
The three-dimensional (3D) SOI correction to the free electronic Hamiltonian due to BIA reads:
%
%
\begin{align}\label{Ham:Dress_3D}
\mathcal{H}_\mathrm{DSOI}^{3D} \propto p_x \left(p_y^2-p_z^2 \right) s_x + p_y &\left(p_z^2-p_x^2 \right) s_y \nonumber \\
&+ p_z \left(p_x^2-p_y^2 \right) s_z  \:.
\end{align}
%
%
In order to achieve the effective Hamiltonian acting on the electrons confined in the 2DEG we have to integrate along the growth direction with the constraint that $\langle p_z \rangle = 0$, while $\langle p_z^2 \rangle \neq 0$ is  a sample dependent constant.
Therefore we have
%
%
\begin{equation}\label{Ham:Dress_2D}
\mathcal{H}_\mathrm{DSOI} = \beta \left(p_y s_y-p_x s_x\right) +\gamma \left(p_x p_y^2 s_x -p_y p_x^2 s_y \right)  
\end{equation}
%
%
where $\gamma$ is a material dependent constant, and $\beta$ depends on $\langle p_z ^2\rangle$. The first term contains the  so called linear DSOI whereas the second one describes the cubic one \cite{Winkler:book}. Usually the latter can be disregarded as $\langle p_z ^2\rangle\ll p_x,p_y $ in a 2DEG.

Due to charging effects, the quantum well confining electrons in two dimensions is never perfectly flat, therefore at the interface between the two differently doped semiconductors, a potential gradient arises $\bm{\nabla} V= \bm{E}$. This effective electric field couples with the electrons motion as 
%
%
\begin{equation}\label{Ham:RSOI_0}
\mathcal{H}_\mathrm{RSOI} \propto \left(\bm{E} \times \bm{p} \right)\cdot \bm{s}\:.
\end{equation}
%
%
Considering a quantum well along the $(001)$ growth direction $\bm{E} = E_z  \bm{z}$, the RSOI is rephrased as:
%
%
\begin{equation}\label{Ham:RSOI}
\mathcal{H}_\mathrm{RSOI}=\frac{\alpha}{\hbar}( \bm{s}\times \bm{p})\cdot \bm{z}=\frac{\alpha}{\hbar}(p_y s_x-p_x s_y)\,,
\end{equation}
%
%
where $\alpha$ depends on the material and on the confining potential.
By simple inspection one can notice that Rashba Hamiltonian \eqref{Ham:RSOI} and linear Dresselhaus SOI \eqref{Ham:Dress_2D} are equivalent under the unitary transformation rotating the spin Pauli matrices $s_x \rightarrow s_y$, $s_y \rightarrow s_x$ and $s_z \rightarrow -s_z$ therefore the spectral properties of electrons subject to RSOI or linear DSOI are exactly the same \cite{Lucignano:2008}. It is important to notice that through a simple mathematical manipulation we can recast RSOI in the form 
$\bm{s}\cdot\bm{B}_\mathrm{eff}(\bm{p})$ where $\bm{B}_\mathrm{eff} \propto (\bm{z} \times\bm{p})$  is an effective magnetic field rotating the carrier spin in the plane of the 2DEG. 

Despite this formal equivalence, the role to RSOI has been investigated with much more attention in the last decades for two main reasons. First, in conventional III-V 2DEGs $\alpha>\beta$. Second, more interestingly, from few decades $\alpha$ can be easily tuned by gating the heterostructure~\cite{silva:1994,silva:1997}  to some hundreds of meV nm. The spin precession length can be expressed as
%
%
\begin{equation*}
\ell_\text{SO} = \frac{\pi}{k_\text{SO}} = \frac{\pi \hbar^2}{\alpha m^*}\,,
\end{equation*}
%
%
\noindent in III-V quantum wells it ranges from hundreds of nm to few $\mu m$, thus comparable with typical dimensions achievable in the fabrication of modern mesoscopic devices.
%
%
\begin{figure}[!t]
\centering
\includegraphics[width=\columnwidth]{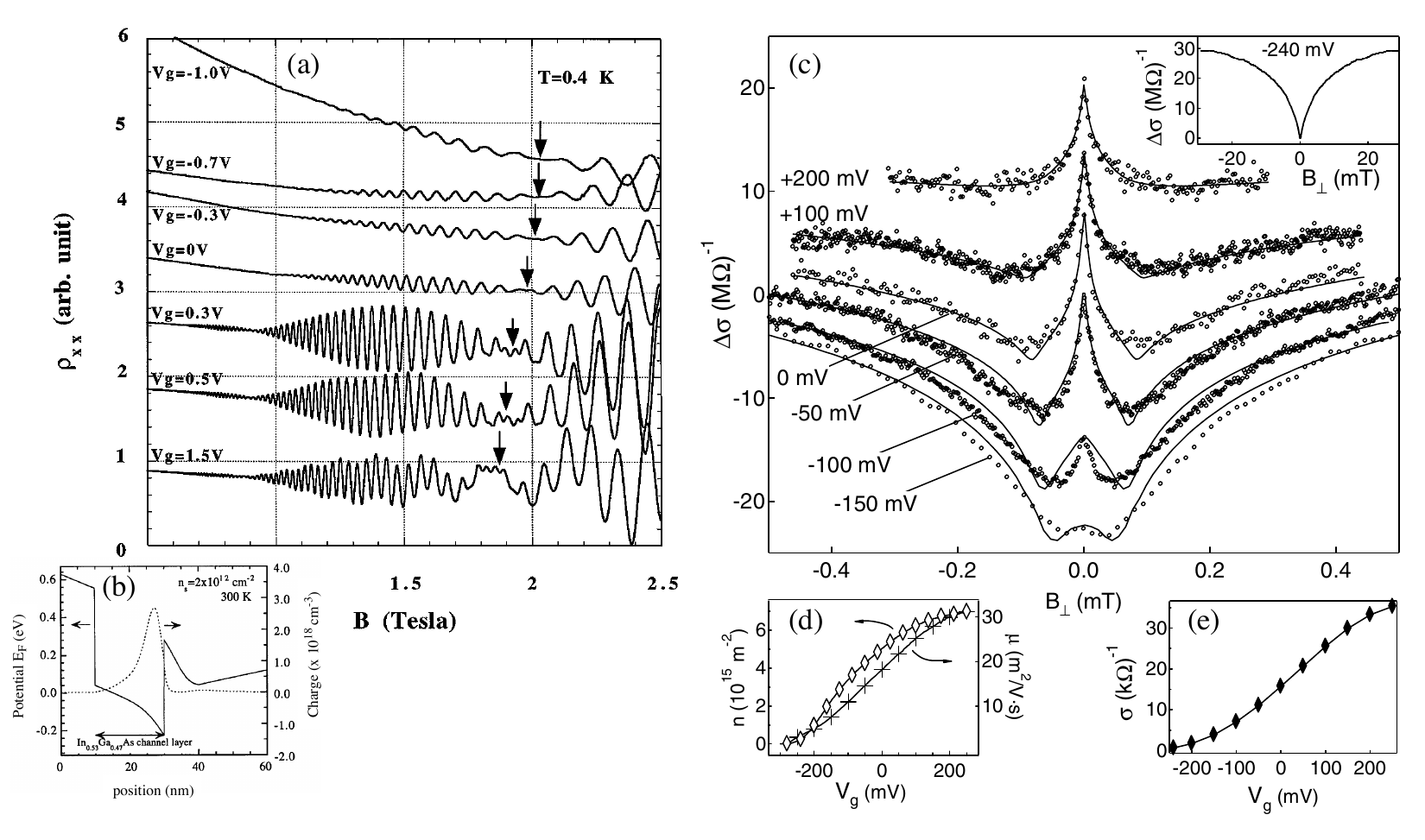}
\caption{\label{alpha:variations} (a) Schubnikov-de Haas oscillations as function of the gate voltages~\cite{nitta:1997}; (b) 
ted conduction band diagram and electron distribution. Panels (a) and (b) courtesy of Ref.~[\onlinecite{nitta:1997}]; (c) Experimental magneto-conductance, $\Delta\sigma=\sigma(b)-\sigma(0)$ (circles), offset for clarity, along with three-parameter fits (solid line) for several gate voltages. Inset: Experimental magneto-conductance data for the most negative gate voltage, showing pure weak localization. (d) Density and mobility as function of $V_\mathrm{g}$, extracted from longitudinal and Hall voltage measurements; (e) Experimental conductivity, showing strong dependence on $V_\mathrm{g}$. Panels (c)-(e) courtesy of Ref.~[\onlinecite{miller:2003}].}
\end{figure}
%
%
The RSOI was first predicted in bulk semiconductors by Rashba~\cite{rashba:1960}. Later Bychkov and Rashba~\cite{bychkov:1984} proposed to use the Hamiltonian \eqref{Ham:RSOI} in order to explain some magnetic properties of semiconductor heterostructures~\cite{Stein:1983,Stormer:1983}. In \fref{alpha:variations}(b) we show the  shape of the quantum well and of the electron density profile showing the position of the 2DEG. 
The lack of inversion symmetry in semiconducting 2DEG heterostructures is at the origin of the RSOI. Therefore a change in the asymmetry can also lead to a change in the strength of the RSOI. For example, by applying a gate voltage is possible to modify the profile of the potential confining the 2DEG with a secondary effect of changing the strength of the RSOI. Thus, in a first approximation the coupling constant $\alpha$ depends on the gate voltage $\alpha\to\alpha(V)$.

A change in the strength of the  $\alpha$, thus,  can produce  a change in the beating pattern of the Schubnikov-de Haas oscillations~\cite{silva:1994,silva:1997} [c.f. \Fref{alpha:variations}(a)]. This effect was first observed by the group of Nitta~\cite{nitta:1997} and after by  other groups~\cite{schapers:1998,grundler:2000}. Another method for probing the variation of the coupling constant is to observe the transition from WL to weak WAL~\cite{Akkermans:2007, miller:2003}\footnote{More details about this  phenomenon will be given in Sec.~\ref{WAL}.}, see for instance  \fref{alpha:variations}(c)-(e).

%
%
\subsection{Quantum wires} 
By applying a further confinement to a 2DEG it  is possible to confine the electrons in a  quasi-1D structure --- a quantum wire. A detailed investigation of the RSOI in InGaAs/InP quantum wires of different width has been put forward by Sch\"apers \emph{et~al.}~[\onlinecite{Schapers:2004,Schaepers:2006}]. Quantum wires  can  also be made out of InAs, GaAs, InSb or other materials~\cite{Nilsson:2009,NadjPerge:2012,EstevezHernandez:2010}. Recently,  quantum wires with sizable RSOI are attracting a lot of interest as they are the fundamental building block for  many proposals  aiming at the observation of Majorana quasi particles in condensed matter physics. In these setups  the interplay between RSOI, superconductivity and  magnetic fields~\cite{oreg:2010,lutchyn:2010} is pivotal.\footnote{Some more analysis of Majorana physics is reported in Sec.~\ref{MFsec}.}

%
%
\subsection{Carbon--based materials}
%
%
\begin{figure}[!ht]
\centering
\includegraphics[width=\columnwidth]{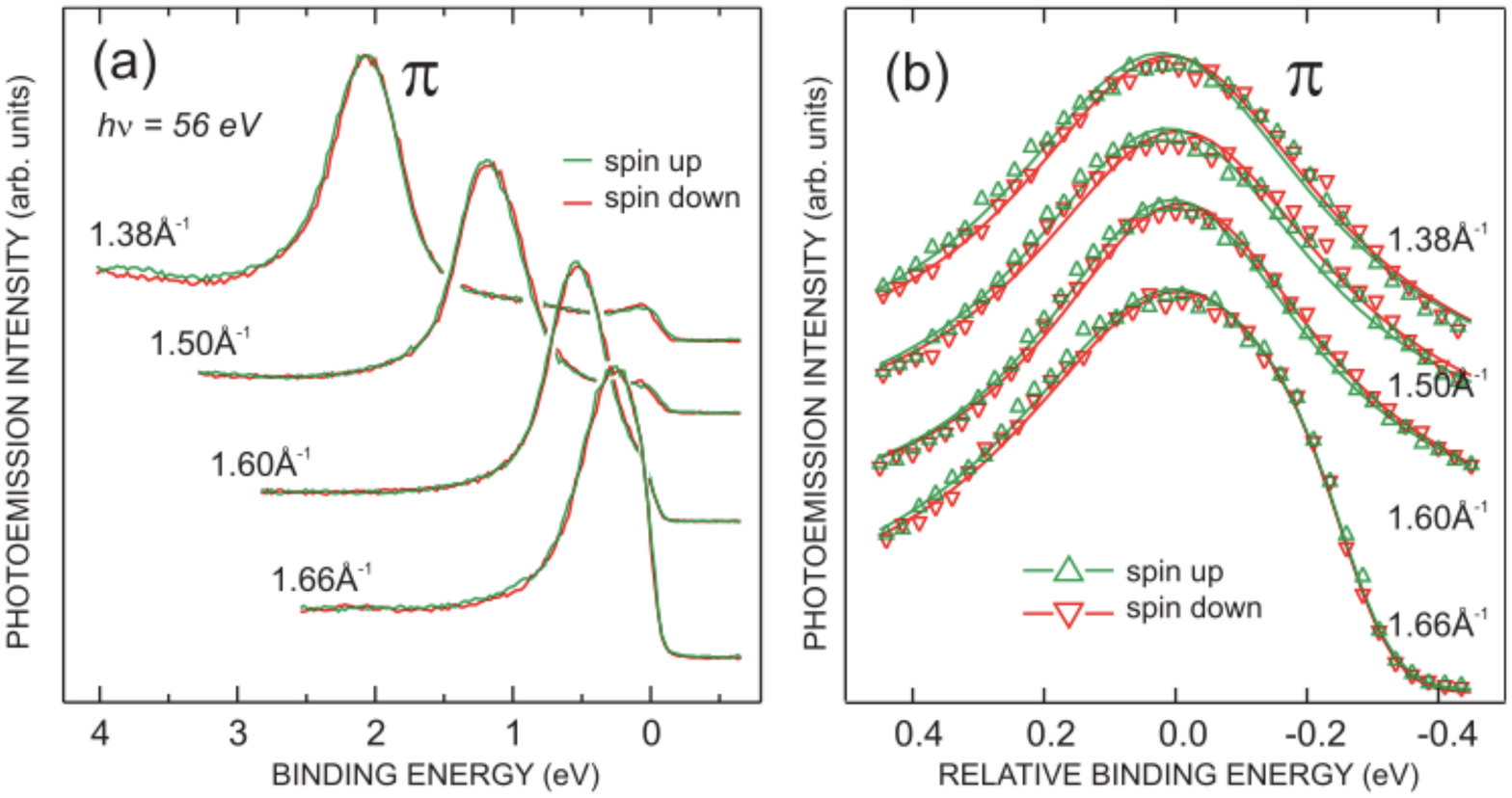}
\caption{\label{fig:RSOI:SLG} (Colours online) Spin-resolved photoemission spectra along $\overline{\Gamma\textrm{K}}$ in the vicinity of the Fermi energy. (a) Overview spectra and (b) $\pi$ states at arbitrary offset. Courtesy of Ref.[\onlinecite{varykhalov:2008}].}
\end{figure}
%
%
The RSOI has been predicted also for single layer graphene (SLG), using simple symmetry arguments~\cite{kane:2005a,kane:2005b}. A more rigorous calculation has been  proposed by Huertas-Hernando \emph{et al.}~\cite{huertas-hernando:2006}, where a modulation via atomic Stark effect and curvature effects were considered. However, due to the small atomic number of the carbon atoms [see equation~\eqref{soi:atom}]  the strength of the coupling $\alpha$ is very small in the SLG structure. There are proposal for overcoming this obstacle and increasing the strength of the SOI, by  coating the graphene surface with $ad$-atoms thus inducing  a  spin-dependent hopping mediated by the surface impurities. This is the case, for example,  of SLG deposited on a Ni substrate and intercalated with Au atoms. Spin-resolved angle resolved photoemission spectroscopy measurements have shown a sizable RSOI of the order of some meV~\cite{varykhalov:2008,Dedkov:2008,Li:2011,marchenko:2012}. Similar proposal have been advanced also for hydrogen impurities~\cite{Neto:2009b}. The modification to the energy spectrum of SLG due to RSOI are shown in \fref{fig:RSOI:SLG}. Large RSOI has been also proposed by using rotating magnetic fields~\cite{Klinovaja:2013}.

The RSOI in SLG is described by a Hamiltonian very similar to \eqref{Ham:RSOI}, the main difference is that  it depends only on the pseudo-spin and not on the momentum operators. To the lowest order in the momentum expansion:
%
%
\begin{equation}\label{soi:slg}
\mathcal{H}_\text{SO}^\text{SLG}= \frac{\lambda}{2}\left(\bm{s}\times\bm{\sigma}\right)_z\,,
\end{equation}
%
%
where $\bm{s}$ are the Pauli matrices describing the electron spin and $\bm{\sigma}$  its pseudo-spin. Here the pseudo-spin is the internal degree of freedom describing the presence of two inequivalent carbon atoms in the honeycomb lattice of SLG~\cite{Neto:2009}.

The RSOI has been predicted also in bi-layer graphene, due to the interaction with a substrate of and an external electric field~\cite{vanGelderen:2010,guinea:2010,Konschuh:2012,Mireles:2012}, however so far it is has not been measured. 

In order to conclude this overview of carbon-based materials with RSOI, relevant effects are predicted also for carbon nanotubes (CNTs)~\cite{Martino:2002,Martino:2005,izumida:2009}. In Refs.~[\onlinecite{kuemmeth:2008,jespersen:2011}]  the RSOI interaction strength is measured in CNT quantum dots.
%
%
\subsection{Topological insulators} 

Topological insulators are band insulators hosting spin polarised edges states within the bulk gap~\cite{Hasan:2010,Qi:2011,Ando:2013}.

\begin{figure}[!t]
\centering
\includegraphics[width=\columnwidth]{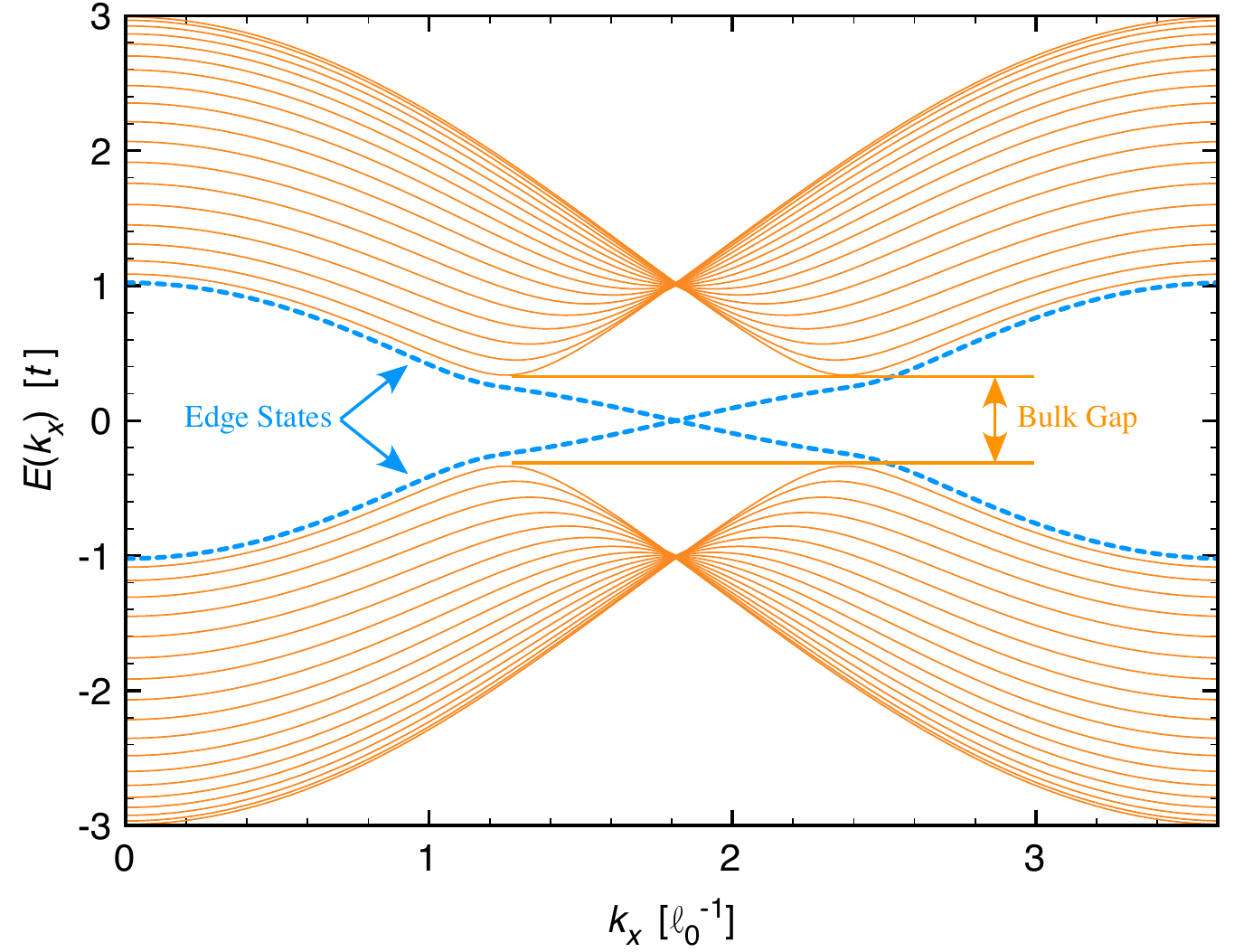}
\caption{(Colours online)\label{ESG} Edge states of the quantum spin-Hall effect in zig-zag graphene nano ribbons, the bulk states are given by the solid-orange lines, the spin-polarised states are the dashed-blue lines.}
\end{figure}

The original idea about this new state of 
matter was put forward by Kane and Mele in two seminal research papers published in 2005~\cite{kane:2005a,kane:2005b}. They showed that a specific type of SOI in SLG can lead to the opening of a gap in the spectrum and that edge states exists within this gap {when considering a ribbon geometry}. 
In \fref{ESG} we can observe the gap opened by the intrinsic SOI~\cite{huertas-hernando:2006} and the edge states within the gap. In two-dimensions, this phenomenon is also known as quantum spin-Hall effect (QSHE). It can be thought as two copies (one per spin) of the  integer quantum Hall effect (IQHE). The main difference is that, contrary to the magnetic field,  SOI does not break time-reserval symmetry.  Thus, here, there are two counter propagating helical edge states opposed to the single chiral edge state of the quantum Hall fluid. As in the case of the IQHE, the edge states  are  characteristics only of the finite size systems. It is also possible to introduce a specific topological numbers named $\mathbb{Z}_2$ that  defines the robustness of these edge states~\cite{kane:2005a}.

Up to now, such phenomenon has not been  observed in SLG, however shortly after the seminal works of Kane and Mele, a  proposal for observing the QSHE in HgTe/CdHgTe quantum wells was put forward by Bernevig \emph{et al.}~[\onlinecite{Bernevig:2006a,Bernevig:2006b}], whose prediction was readily confirmed  by  K\"onig \emph{et al.}~[\onlinecite{Konig:2006}]. Since than, there is a huge quest for discovering new 2D TIs. One of the most interesting is represented by InAs/GaSb quantum well~\cite{liu:2008,knez:2011} which is very attractive as it is achievable using the standard and, very well known, III-V semiconductor technology. 

Contrary to the case of IQHE, the paradigm of the QSHE can be extended to 3D. This work has been mainly carried out by Fu and Kane~[\onlinecite{Fu:2007a,Fu:2007b}]. Since their seminal works, a race for discovering new 3D TIs is open. Recent overviews on these materials can be found in one of the several review articles that has been published on the topic~\cite{Hasan:2010,Qi:2011,Ando:2013}. Among them the paper by  Ando~\cite{Ando:2013} contains an interesting  table summarising  all known 2D and 3D TI discovered up to 2013. 

\begin{figure}
\centering
\includegraphics[width=\columnwidth]{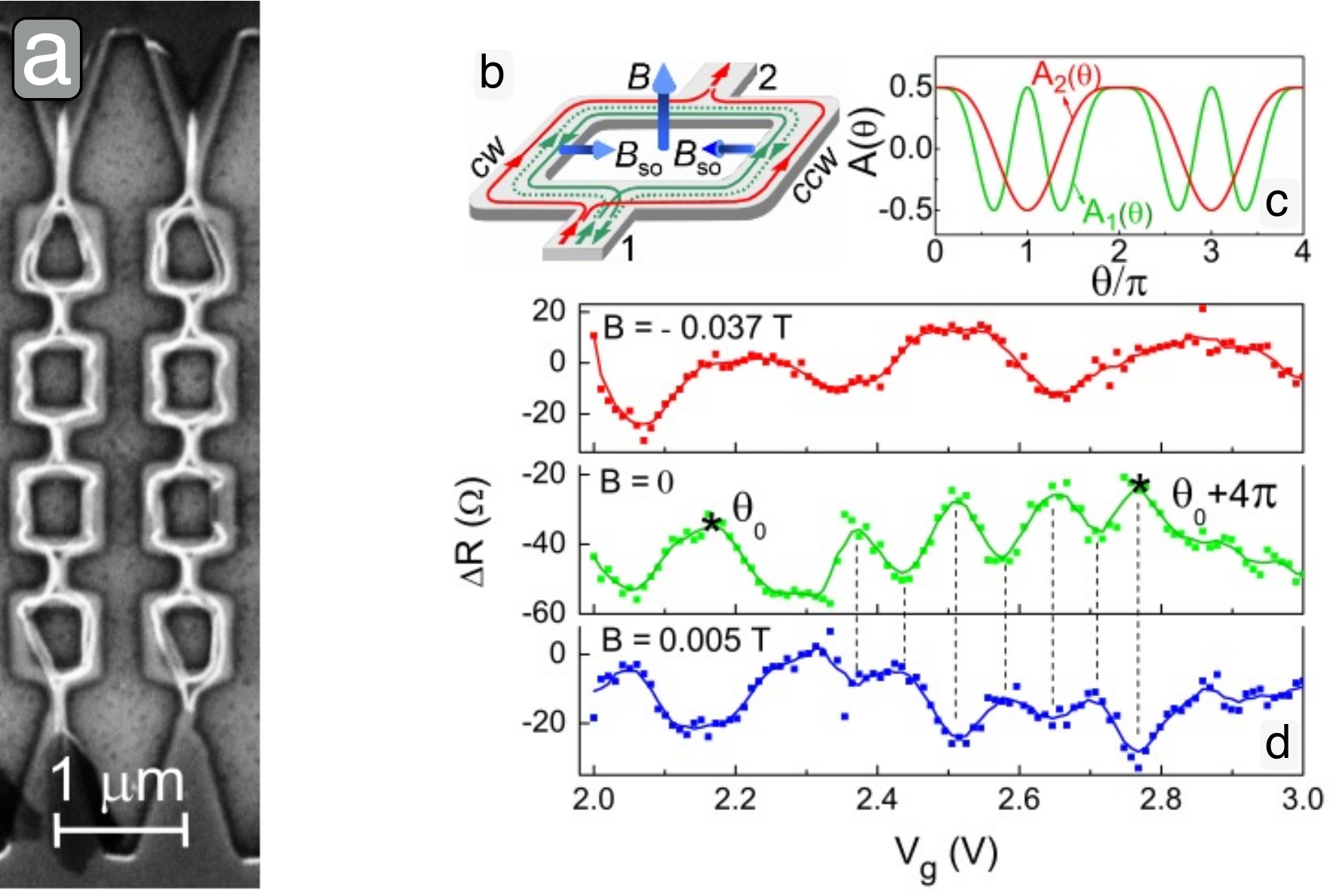}
\caption{\label{Bi2Se3}(Colours online) (a): SEM image of the square-ring device realised in Bi$_2$Se$_3$; (b): Illustration of the AB effect interference (red trajectories) and  AAS interference (solid and dotted green loops) for charges in a square ring. The existence of RSOI creates an effective magnetic field $B_\text{SO}$ pointing towards/ or from the center of the ring for counterclockwise (CCW)/clockwise (CW) propagation modes, which influences spin precession and generates an AC phase in addition to the AB and AAS phases. 
(c) $A_1(\theta)$ and $A_2(\theta)$ as a function of $\theta$. Here $\theta=2\alpha m^* L / \hbar^2$ is spin precession angle over a distance $L$ and $A_1(\theta)=(\cos^4\theta+4\cos\theta\sin^2\theta+\cos2\theta)/4$ and $A_2(\theta)=(\sin^2\theta + \cos2\theta)/4$ are the modulation of the wave function as a function of $\theta$ (c.f. Ref.~\cite{Qu:2011} for more details).
The former varies roughly at twice frequency of the latter. (d): $\Delta R-V_\mathrm{g}$ curves at fixed magnetic fields marked by the lines and arrows of corresponding colours. The spin precession angle is modulated by $4\pi$ by varying $V_\mathrm{g}$ from the interval $2.16$ to $2.77$ V marked by the two stars. The dashed lines help illustrating the opposite phases between the green and blue curves in the AAS region.  Courtesy of Ref.~[\onlinecite{Qu:2011}].}
\end{figure}

Most of the research work is devoted to analyse the spectral properties of such material---mainly via angle-resolved photoemission spectroscopy. Only few experiments  show the transport properties of this new class of materials. Mainly because  most of the 3D TI are not proper insulators due to a non negligible current flowing through impurity states~\cite{Bardarson:2013}.
%
%
%
%
A very interesting set of transport experiments has been performed in films of Bi$_{2}$Se$_{3}$~\cite{Qu:2011} prepared with a special two-dimensional geometry (c.f.~\fref{Bi2Se3}). In this work the authors show a sizable tunability of RSOI in 3D TI. The effects of this tunability are clearly observed in the  interference pattern of the Aharonov-Bohm (AB), the Altshuler-Aronov-Spivak (AAS) and the Aharonov-Casher (AC) effects~\cite{AltlandAlexander:2010,Akkermans:2007,Blanter:2013}. This is very interesting because it would allow to realise novel functionalities such as possible non-Abelian operations on spins.
%
%
%
%
%
%
\subsection{Weyl semimetals} 

Weyl semimetals also known as ``topological semimetals'' are  zero gap semiconductors with a Fermi level very close to the centre of the gap that is at zero energy as in the case of SLG and TIs. The  effective Hamiltonian describing the low energy states close to the Fermi level in WSs is represented by a $2\times2$ Hamiltonian that is linear in the momentum --- as for TI and SLG --- but it is an actual 3D model, thus containing all the three Pauli matrices. The simplest model Hamiltonian in $k\,p$ approximation reads:
%
%
\begin{equation}\label{WS:Ham}
\mathcal{H}_\pm= \pm v_\mathrm{F}\bm{p}\cdot\bm{s}\,,
\end{equation}
%
%
where $v_\mathrm{F}$ is the Fermi velocity, and $\bm{p}$ and $\bm{s}$ are the vectors of the momentum and of the Pauli matrices, respectively. By inspection of equation \ref{WS:Ham} we evince  that a generic perturbation cannot gap the energy spectrum. Its energy spectrum reads $E_\pm(p)= \pm v_\mathrm{F}|\bm{p}|$.  This is usually correct for an even number of \emph{nodal} points where the expansion \eqref{WS:Ham} is allowed. At each node we can associate a chirality, this measures the relative handedness of the three momenta and the Pauli matrices associated in the Weyl equation --- the chirality $\pm1$ for the Hamiltonians $\mathcal{H}_\pm$, can also be thought as a source of Berry curvature carried by the WS nodes. This is a general property of Weyl fermions realised in band structures: their net chirality must in fact cancel\footnote{This can be understood in terms of the fermion doubling theorem~\cite{Nielsen:1981}.}. 
Note that it is usually assumed that bands are individually nondegenerate. This requires that either the time-reversal symmetry or the inversion symmetry (parity) is broken. In order to realise the minimal case of just a pair of opposite chirality Weyl nodes, time-reversal symmetry must be broken~\cite{Wan:2011}.

The idea of WS was first theoretically proposed by Murakami in 2007~[\onlinecite{Murakami:2007}] and later it was further elaborated by Wan \emph{et~al.} in 2011~[\onlinecite{Wan:2011}]. Based on \emph{ab}-initio band calculations, it was proposed that pyrochlore iridates, such as Y$_2$Ir$_2$O$_7$, in the antiferromagnetic phase, may realise such a WS~[\onlinecite{Wan:2011,Yang:2010}]. A recent review on the spectral properties of this type of system is reported by Vafek and Vishwanath~[\onlinecite{Vafek:2014}] and on the transport properties by Hosur and Qi~[\onlinecite{Hosur:2013}].

\section{Spin transport in RSOI material}\label{transport}

Before describing the role of RSOI in the spin-dependent transport, it is important to understand how it modifies the spectral properties of  a free electron in a 2DEG.

\subsection{General properties of the RSOI in semiconductors}\label{RSOI2DEG}

\begin{figure}
	\centering
	\includegraphics[width=\columnwidth]{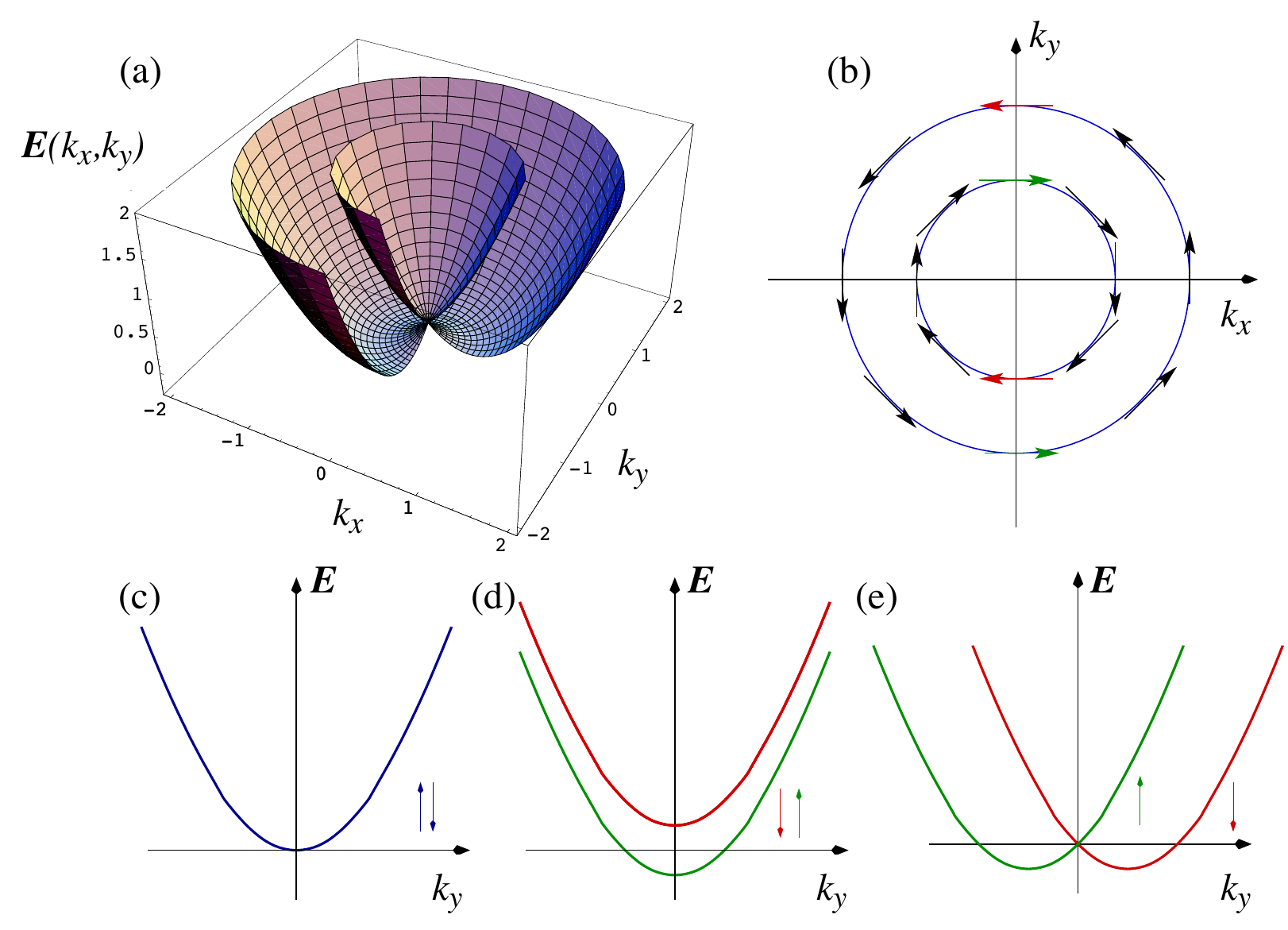}
	\caption{(Colours online) Properties of the Rashba energy spectrum. (a) Portion of the energy spectrum of the Hamiltonian \eqref{rashba2DEG}. (b) The Fermi contours relative to the Hamiltonian~\eqref{rashba2DEG}, the spin states are shown as well. (c) Section of the energy spectrum for a free electron. (d) Section of the energy spectrum for an electron in presence of a magnetic field, \emph{e.g.}~\textit{Zeeman} splitting. (e) Section of the energy spectrum for an electron in presence of RSOI.\label{spectrum2d}}
\end{figure}

We consider a 2DEG in the  $(x,y)$--plane in the presence of the RSOI \eqref{Ham:RSOI} and with effective electron mass $m^*$. The full Hamiltonian of the system is
%
%
\begin{equation}\label{rashba2DEG}
	\mathcal{H}_0 = \displaystyle\frac{\bm{p}^2}{2m^*} +
	\frac{\alpha}{\hbar} (\bm{s}\times\bm{p} )_{z} .
\end{equation}
%
%
The eigenvalues read 
%
%
\begin{equation}\label{spectrumRashba}
\mathcal{E}_\pm(\bm{k}) = \displaystyle\frac{\hbar^2 k^2}{2m^*} \pm \alpha
k = \displaystyle\frac{\hbar^2}{2m^*} \left(k \pm k_{\mathrm{SO}}
\right)^2 - \Delta_{\mathrm{SO}},
\end{equation}
%
%
where $k=|\bm{k}|=\sqrt{k_x^2+k_y^2}$ is the modulus of the electron momentum,
$k_{\mathrm{SO}} = \frac{\alpha m^*}{\hbar^2}$ is the RSOI coupling constant with the dimension of a momentum and $\Delta_{\mathrm{SO}} = \left(\frac{\alpha m^*}{\hbar}\right)^2$. Usually the last term of (\ref{spectrumRashba}) is neglected because it is second order in $\alpha$. Moreover, even if included, it would lead to a rigid shift of the bands, thus renormalising the chemical potential $\mu$. The eigenvectors of the Hamiltonian
\eqref{rashba2DEG} relative to the spectrum \eqref{spectrumRashba} are plane waves:
%
%
\begin{equation}\label{eigenvectorOne}
\Psi_\pm(\bm{r}) =\displaystyle 
\frac{\rme^{\rmi\bm{k}\cdot\bm{r}}}{\sqrt{2}} \left( \begin{array}{cc} 1 \\ \pm\rmi\rme^{-\rmi\theta}\end{array}\right),  
\end{equation}
%
%
where $\theta=\arctan(k_y/k_x)$ is the polar angle between the momentum vector and the $k_x$ direction. It is important to note that the spin states \eqref{eigenvectorOne} are always perpendicular to the motion direction. In fact, if an electron moves along $x$ direction the spinor part of the eigenvectors become $\left(1;\rmi\right)$ and $\left({1};-\rmi\right)$ that is the spin up and spin down are along the $y$ direction. By contrast,
 if the electron moves along the $y$ direction, the eigenvectors become $\left(1; 1\right)$ and $\left(1; -1\right)$ that is the spin up and spin down
state in the $x$ direction [see \fref{spectrum2d}(b)].
 
In \fref{spectrum2d}(c)-(e) we report the $k_y$-section of the energy {spectrum \emph{vs} the momentum} for a 2DEG in different physical situations.  In \ref{spectrum2d}(c) we considered a free particle in a 2DEG. In this case the spectrum is twofold degenerate in spin. In the presence of a magnetic field $\bm{B}$  [see panel \ref{spectrum2d}(d)], the spin degeneracy is lifted out by the \textit{Zeeman} effect and the gap separating spin up and spin down is equal to $2 g^* \mu_\mathrm{B} B$, where $g^*$ is the effective gyromagnetic ratio and $\mu_\mathrm{B}$ is the Bohr's magneton. When the RSOI  is present \ref{spectrum2d}(e), the spin degeneracy is lifted out except for $k_y=0$. In this situation the degeneracy is removed without the opening of gaps.
%
%
%
%
%
The semiclassical particle velocities are given by 
%
%
\begin{equation}\label{velocity}
\bm{v}_\pm (\bm{k}) = \displaystyle\frac{1}{\hbar}\frac{\partial \mathcal{E}_\pm
(\bm{k})}{\partial \bm{k}} = \frac{\hbar \bm{k}}{m^*} \pm \alpha\bm{k}=\frac{\hbar}{m^*}(k\pm k_\mathrm{SO})\bm{k}.
\end{equation}
%
%
If we consider the quantum-mechanical velocity operator
%
%
\begin{equation}
\dot{\bm{r}} = \displaystyle\frac{\rmi}{\hbar} \left[ \mathcal{H}_0 ,
\bm{r} \right]
\end{equation}
%
%
and the expression \eqref{eigenvectorOne} for the eigenstates, it is straightforward to show that its matrix elements are given by
%
%
\begin{equation}
\langle \Psi_\pm (\bm{k})|~ \dot{\bm{r}} ~| \Psi_\pm (\bm{k}')
\rangle = \delta(\bm{k}-\bm{k}')~ \bm{v}_\pm (\bm{k})\,.
\end{equation}
%
%
That is the semiclassical velocities $\bm{v}_\pm (\bm{k})$ are, as
usual, the diagonal elements of the velocity operator. In the presence of the RSOI, the velocity operator is not simply the momentum divided by the effective mass as  for free electrons and the Fermi surface splits into two surfaces shown in \fref{spectrum2d}(b).
One can easily observe that counter propagating electrons have opposite spins. Spin and momentum are locked to each other. This will have importance consequences in the following. Parametrising the wave vectors as $\bm{k} = k ~(\cos \varphi,\sin\varphi)$ the two Fermi surfaces are described by the following equation:
%
%
\begin{equation}\label{kfermi}
 k_{\pm}^\mathrm{F} (\mathcal{E}_\mathrm{F}) = \mp
\displaystyle\frac{\alpha ~m^*}{\hbar^2} + \sqrt{\left(\frac{\alpha
~m^*}{\hbar^2}\right)^2 + \frac{2m^*}{\hbar^2} \mathcal{E}_\mathrm{F}}\,. 
\end{equation}
%
%
Here the double sign corresponds to the two dispersion branches
(\ref{spectrumRashba}).

\subsection{Rashba SOI in quantum wires: subbands hybridisation} \label{sub:Hyb}
By further confining a 2DEG along one direction (we choose $y$ in the following) we can realise a quantum wire. In this case we cannot solve analytically the system Hamiltonian,  as we have done  for the simple case of a 2DEG. Different theoretical models have been used to describe confinement effects. A very convenient representation consists in assuming a transversal confining potential in the $y$ directions, and let the electrons propagate along the  the $x$ direction. 
Such choice is particularly  convenient when expressing the matrix elements of the SOI~\cite{Moroz:1999,Governale:2002,perroni:2007}. 
The Hamiltonian reads:
%
%
\begin{equation}\label{hamiltonian_wire}
H=\mathcal{H}_\parallel+\mathcal{H}_\perp+\mathcal{H}_\mathrm{mix}\,,
\end{equation}
%
%
with the following terms:
%
%
\begin{subequations}
\begin{eqnarray}
\mathcal{H}_\parallel& = &\frac{p_x^2}{2m^*}-\frac{\hbar k_\mathrm{SO} }{m^*} \sigma_y p_x \label{QW1}\\
\mathcal{H}_\perp & = &\frac{p_y^2}{2m^*} + V(y) \label{QW2} \\
\mathcal{H}_\mathrm{mix} & = & \frac{\alpha}{\hbar} \sigma_x p_y\,. \label{QW3}
\end{eqnarray}
\end{subequations}
%
%
Where $V(y)$ is a infinite well potential of width $W$.
Neglecting $\mathcal{H}_\mathrm{mix}$, the terms in equations \eqref{QW1} and \eqref{QW2} do commute, therefore the eigenvalues and eigenvectors can be easily expressed as: 
%
%
\begin{align}
E_{\parallel +\perp}(n,\sigma_y,k_x) &= \frac{\hbar ^2} {2 m^*}\left(k_x^2+\frac{\pi^2 n^2}{W^2}\right) - \frac{\hbar^2 k_\mathrm{SO} k_x}{m^*}\langle \sigma_y\rangle\\
\psi_{n \sigma_y k_x} (x,y)& = \phi_n(y) \psi_{k_x}(x)|\sigma_y\rangle
\end{align}
%
%
%
where $\phi_n(y)$ are  the $n$-th eigenfunctions of the potential $V(y)$ that are a either a sine or a cosine functions and $\psi_{k_x}(x)$ simple plane waves shifted by the RSOI, {here $\langle \sigma_y \rangle$ can get the values $\pm1$}. The term $\mathcal{H}_\mathrm{mix}$ induces mixing between these states and gives rise to a deformation of the electronic bands and to anti-crossings in the energy spectrum.  

\begin{figure}[!t]
\centering
\includegraphics[width=0.7\columnwidth]{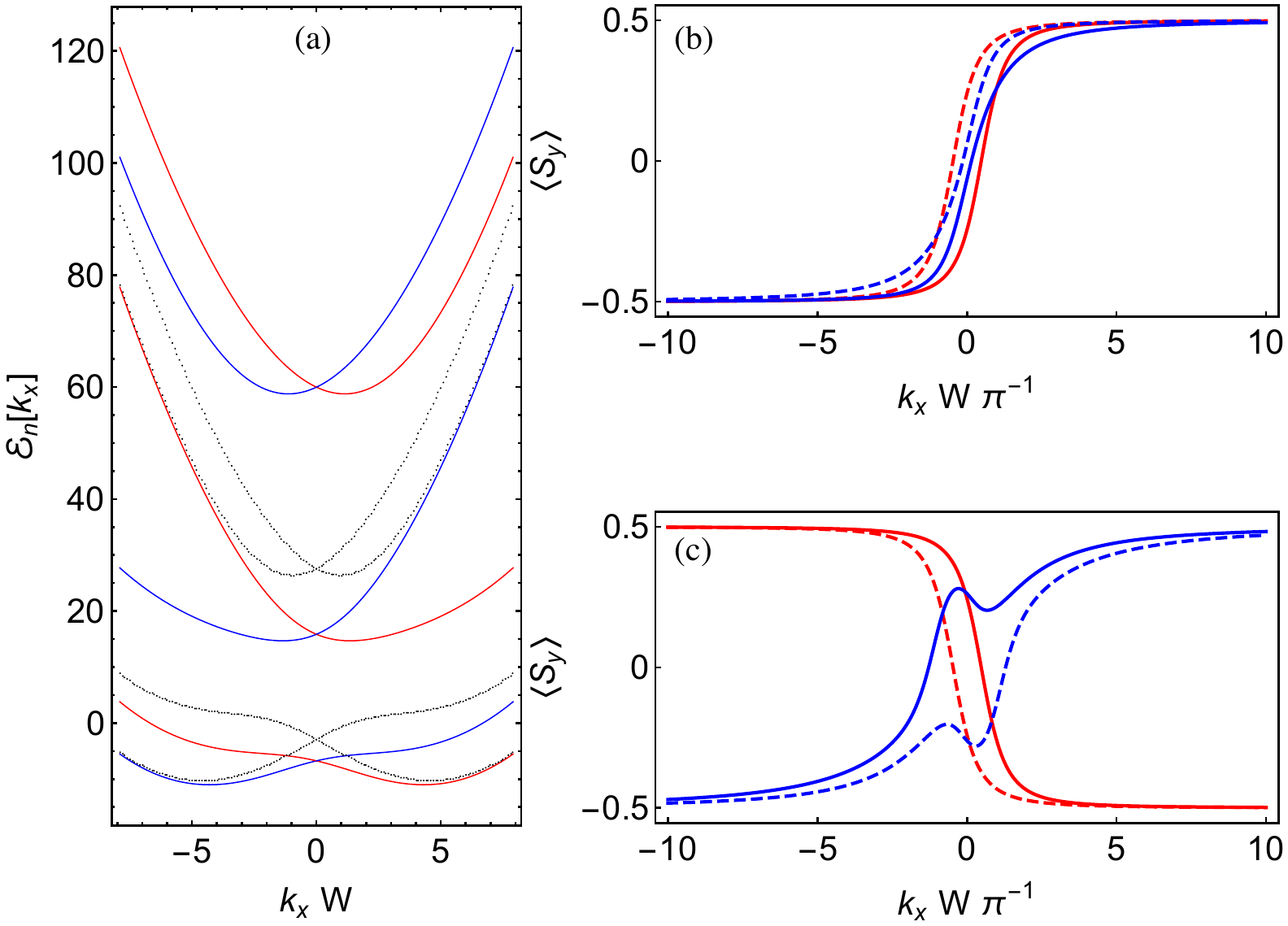}
\caption{\label{wire}(Colours online) (a) Band structure of quantum wires with square well potential along the confining direction. The coloured lines refer to the first three sub-bands in the case of a wire with three modes whereas the dotted-black lines refer to the same wire but only considering two modes. (b) Spin polarisation $\langle S_y\rangle$ as a function of the transversal momentum for the first sub-band in the case of a two band model (red-solid and red-dashed lines). (c) Spin polarisation $\langle S_y\rangle$ as a function of the transversal momentum for the second sub-band in the case of a two band model (red-solid and red-dashed lines).  Both in panel (b) and (c) the spin polarisation is evaluated using a truncated Hilbert space containing 50 modes (blue-solid and blue-dashed lines).}
\end{figure}

Anti-crossings occur between sub-bands corresponding to transverse eigenstates with different $n$ [\fref{wire}(a)]. 
%
%
%
%
While simple one dimensional (1D) models predict a rigid spin-momentum locking with \emph{in}--plane spin perpendicular to the momentum along the wire direction, multi-band models predict that only  electrons with momentum far away from anti-crossings have the spin essentially perpendicular to the momentum, \emph{i.e.} $\langle S_y \rangle \sim \pm \frac{1}{2}$ and $\bm{k}\sim k_x \bm x$. This is well shown in \fref{wire}(b), where we can see the expectation value of $\langle S_y\rangle$ as a function of the longitudinal momentum $k_x$ --- it changes form $-\frac{1}{2}$ to $\frac{1}{2}$ for $k_x$ changing from negative to positive values. The same picture applies to the other sub-bands. Here we shall focus on  the role played by the number of sub-bands used for evaluating the system properties. The Hamiltonian \eqref{hamiltonian_wire} cannot be diagonalized exactly, but a partial analytical/numerical solution can be obtained by truncating the Hilbert space associated to the modes produced by  $V(y)$.  
Diagonalizing the Hamiltonian~\eqref{hamiltonian_wire} for $N\gg1$ modes produces a result that is substantially correct for the lowest  $N-1$ modes but not for the mode $N$. This is observed, \emph{e.g.}, in the spectral properties and in the spin polarisation. The spectrum is shown in \fref{wire}(a),  we have in colours-solid lines for $N=3$ and in black-dashed lines for $N=2$. We see that the first sub-bands in both cases ($N=3$ and $N=2$) are quite similar apart from a negative energy shift~\cite{perroni:2007}. However, the situation is very different when we consider the second sub-bands, these greatly differ in the two cases --- due to the presence of the third mode for the case $N=3$. Considering the spin polarisation the effect is more visible. In \fref{wire}(b)  the spin polarisations for the first sub-bands for $N=2$ (red lines) and $N=50$ (blue lines) are shown, the behaviours are similar in both cases but around $k_x\sim0$. The  error produced by Hilbert space truncation is shown in \ref{wire}(c), here it is shown the spin polarisation for the second sub-bands (red lines), for $N=2$. We observe a mirroring of the results reported in \ref{wire}(b). However, the most correct result with $N=50$ (blue lines) contradicts completely $N=2$.

The  message here is that when solving a problem of quantum transport in a confined geometry a sufficient number of sub-bands should be taken into account in order to avoid systematic errors due to the truncation of the Hilbert space~\cite{Moroz:1999,Governale:2002,perroni:2007}. The same conclusions are obtained by changing the shape of the confining potential $V(y)$ from square potential one to harmonic oscillator one~\cite{Moroz:1999,Governale:2002} or by introducing a further periodic modulation along the longitudinal direction $x$~\cite{Smirnov:2007}. The sub-band hybridisation that we have shown before can give rise to dip into the conductance of quantum wires with RSOI as shown in several works both analytically~\cite{Moroz:1999,perroni:2007} and numerically~\cite{Sanchez:2006} and can also be interpreted in terms of a Landau-Zener transition when analysing the conductance of a quantum point contact~\cite{EtoMikio:2013}.

\subsection{General properties of the Rashba SOI  in graphene}

If we restrict ourself to the case of a single valley approximation, the effects of RSOI on SLG via Hamiltonian \eqref{soi:slg} change with respect to what we have seen in the case of a 2DEG in Sec.~\ref{RSOI2DEG}. The complete Hamiltonian for the SLG with RSOI reads:
%
%
\begin{align}\label{SLG:plus:RSOI}
\mathcal{H}_{\bf K} & = v_{\mathrm{F}} \bm{\sigma} \cdot \bm{p} + \frac{\alpha}{2}\left(\bm{s}\times\bm{\sigma}\right)_z\nonumber \\
& =v_{\mathrm{F}} \left(\sigma_x p_x+\sigma_y p_y\right)+ \frac{\alpha}{2}\left(s_x\sigma_y-s_x\sigma_y\right)\,.
\end{align}
%
%
The energy eigenstates are plane waves  $\psi \sim \Phi(\bm{k})\mathrm{e}^{\rmi \bm{k}\cdot\bm{r}}$ with $\Phi$ a four-component spinor and
eigenvalues given by ($v_\text{F}=\hbar = 1$)
%
%
\begin{equation}\label{eq:so:spectrum}
\mathcal{E}_{\pm,\epsilon}(\bm{k}) =\pm\frac{\alpha}{2} +\epsilon \sqrt{k_x^2+k_y^2 + 
\frac{\alpha^2}{4}},
\end{equation}
%
%
where index $\epsilon=\pm$  specifies the particle/hole branches 
of the spectrum. 
%
%
\begin{figure}[!b]
\centering
\includegraphics[width=0.6\columnwidth]{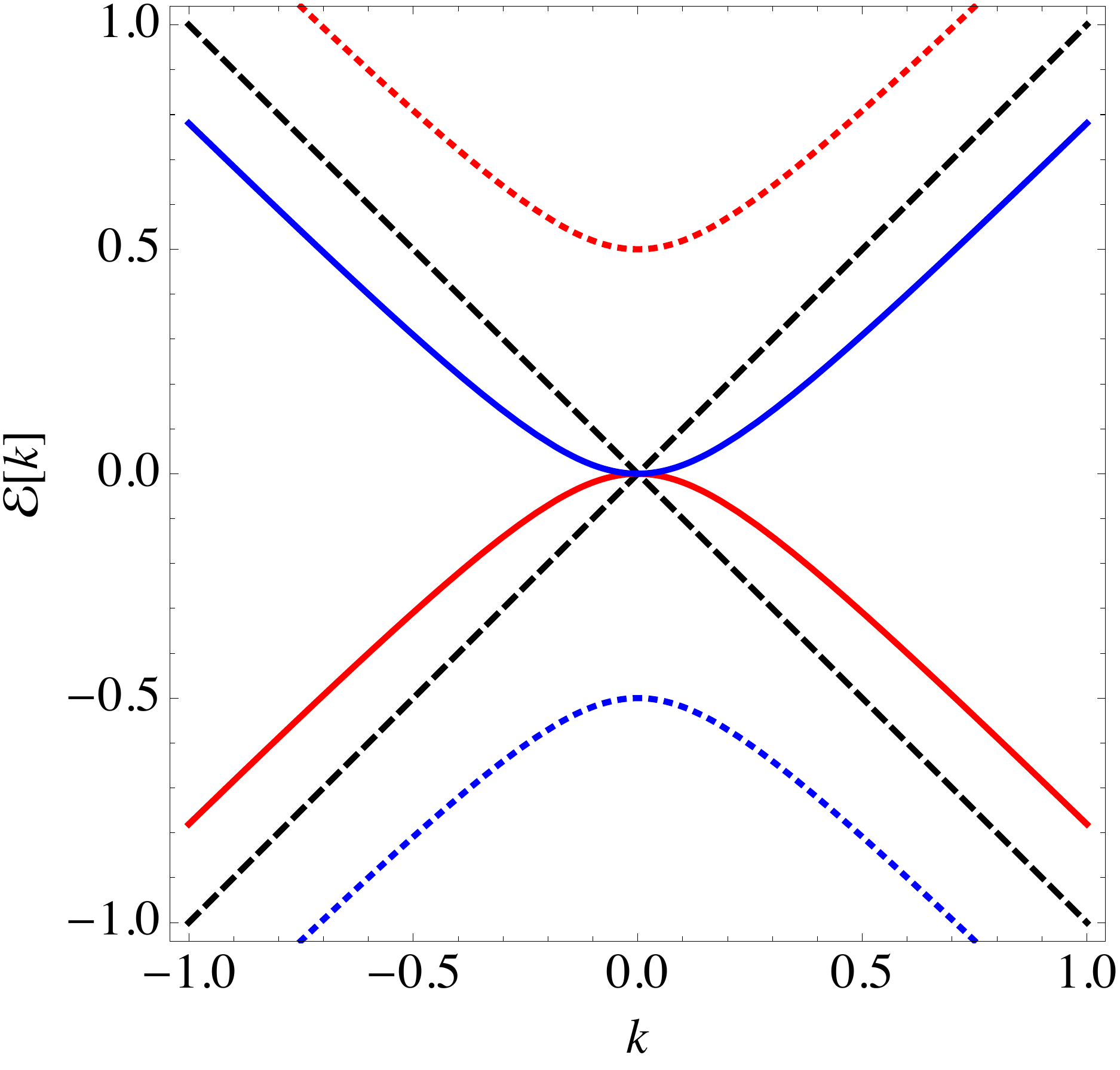}
\caption{\label{spec:HCL}(Colours online) Energy spectrum of graphene for $k_y=0$ as a function of $k$ along the $x$ axis. The dashed lines correspond to the linear dispersion for $\alpha=0$, the solid and the dotted lines to the case of finite RSOI. The lines with the same colours correspond to the same spin state.}
\end{figure}
%
%
The energy dispersion, as a function of $k_x$, at fixed $k_y=0$, is illustrated in \fref{spec:HCL}   with and without RSOI. As we can see here, the RSOI  gives rise to a finite curvature of the linear energy spectrum and lifts the spin degeneracy. Two opposite spin bands get gapped while the other two are still degenerate at $\bm{k}\sim 0$. However, contrary to the case of 2DEGs, the energy spectrum is not shifted along the momentum axis as in equation \eqref{spectrumRashba}. This is related to the fact that the RSOI Hamiltonian does not depend on the electron momentum but only on its pseudo-spin, to the lowest order in the momentum expansion. {Nevertheless}, this approximation no longer holds  if we expand  RSOI Hamiltonian to higher orders, as they  explicitly depend on the electron momentum. Their main effect, for strong RSOI, is to produce the appearing of extra Dirac cones  (tridiagonal wrapping Ref.~[\onlinecite{Zarea:2009,Rakyta:2010,Lenz:2013}]).

The eigenstates of the Hamiltonian \eqref{SLG:plus:RSOI},  are expressed by  the spinors $\Phi_{\pm,\epsilon}({\bm k})$:
\begin{widetext}
%
%
\begin{eqnarray}
\Phi^\mathrm{T}_{\pm,\epsilon} (\bm{k}) =&
\frac{1}{2\sqrt{\cosh \theta_\pm}} 
\times(\rme^{-\rmi \phi-\epsilon \theta_\pm/2},
\epsilon \mathrm{e}^{\epsilon\theta_\pm/2},\pm\rmi\epsilon \mathrm{e}^{\epsilon\theta_\pm/2},\pm\rmi\mathrm{e}^{\rmi\phi-\epsilon\theta_\pm/2}), \nonumber
\end{eqnarray}
%
%
\end{widetext}
where ${\square}^\mathrm{T}$ denotes transposition and
%
%
\begin{eqnarray}
&\sinh \theta_\pm = \pm\frac{\alpha}{2 k},\\
& \mathrm{e}^{\rmi\phi} = \frac{k_x+\rmi k_y}{k}, \label{phi}
\end{eqnarray}
%
%

with $k=\sqrt{k_x^2+k_y^2}$.  The spin operator components are expressed as $S_j=\frac{1}{2}s_j\otimes \mathbb{I}_2$. Their expectation values on the eigenstates $\Phi_{\pm,\epsilon}$ read
%
%
\begin{subequations}
\begin{eqnarray}\label{eq:spin:state}
\langle S_x \rangle &=& -\frac{\epsilon\sin \phi}{2\cosh \theta_\pm},  \\
\langle S_y \rangle &=& \frac{\epsilon\cos \phi}{2\cosh \theta_\pm}, \\
\langle S_z \rangle &=& 0,
\end{eqnarray}
\end{subequations}
%
%
which shows that the product {$\pm\epsilon$} coincides with the sign of the expectation value of the spin projection along the \emph{in}--plane direction perpendicular to the direction of propagation. For vanishing  RSOI, the eigenstates $\Phi_{\pm,\epsilon}$ reduce to linear combinations of eigenstates of $S_z$.
Similarly, the expectation value of the pseudo-spin operator $\bm{\sigma}$ is given by
%
%
\begin{subequations}
\begin{eqnarray}
\langle \sigma_x \rangle &= \frac{\epsilon \cos \phi}{\cosh \theta_\pm}, \label{psx}\\
\langle \sigma_y \rangle &= \frac{\epsilon \sin \phi}{\cosh \theta_\pm}. \label{psy}
\end{eqnarray}
\end{subequations}
%
%
Since the RSOI in SLG does not depend on momentum, the velocity operator still coincides with the pseudo-spin operator:  ${\bm v}=\dot{\bm r} = \rmi [\mathcal{H},{\bm r} ]={\bm{\sigma}}$. Thus,  the velocity expectation value in the state $\Phi_{\pm,\epsilon}$ is given by equations \eqref{psx} and \eqref{psy}. Alternatively, it can be obtained from the energy dispersion as
%
%
\begin{eqnarray}
\bi{v}_{\epsilon} = \nabla_{\bi{k}} \mathcal{E}_{\epsilon} =  \frac{\epsilon \bi{k}}{\sqrt{k^2 +\frac{\alpha^2}{4}}}\,. \label{eq:so:velocityx}
\end{eqnarray}
%
%
If we considered also the intrinsic SOI, the effective Fermi velocity would acquire a more complex dependence from the RSOI~\cite{Lenz:2013,Bercioux:2010}.

Sub-band hybridisation due to RSOI  also occurs in graphene nano-ribbons. However, its manifestation depends on the type of boundary~\cite{Neto:2009}: zig-zag~\cite{Zarea:2009} or armchair~\cite{Lenz:2013}. The most peculiar effects are observed in the case latter. Here in fact, the RSOI couples not only different modes but introduces also a finite coupling inside each sub-band. It is due to the presence of  two inequivalent carbon atoms in the unit cell of a SLG nano ribbon and is completely different from the case of a quantum wire with RSOI of the previous section~\cite{Lenz:2013}. 

\subsection{The spin double refraction}

The physics of an interface between a 2DEG with and without SOI is very similar to that of the optical birefringence~\cite{Born:1999} and is a direct consequence of the two possible Fermi velocities for the two modes \eqref{velocity}. In the following we consider a 2DEG in the $(x,y)$--plane characterised by an effective electron mass $m^*$ and an interface along the $y$ direction separating a region without SOI (N-region) and a region with it (SO region). Elastic scattering at the interface allows for conservation of the total energy and of the momentum parallel to the interface --- $k_y$ in our case. As we have seen in the previous section, the Fermi surface for an energy $\mathcal{E}_\text{F}$ is constituted of two circles with radius~\eqref{kfermi}. An incoming particle from the N region is characterised by the momentum $ k^\text{N}$ and the {incidence} angle $\phi=\arctan\left(\frac{k_y^\text{N}}{k_x^\text{N}}\right)$. Momentum conservation  implies that 
%
%
\begin{equation}\label{kinematicEQ}
k^\text{SO}_+\sin\theta_+=k^\text{SO}_-\sin\theta_-=k^\text{N}\sin\phi\,,
\end{equation}
%
%
where we have introduced $k_\pm^\text{SO}=k_{\pm}^\mathrm{F} (\mathcal{E}_\mathrm{F})$ {and $\theta_\pm$ are the propagation angles in the SO region for the two different modes}. An incoming particle from the normal region propagates into the SO region along two different modes with indices $\pm$. In Refs.~[\onlinecite{perroni:2007,marigliano:2003,khodas:2004,marigliano:2004}]  this phenomenon is named \emph{spin-double refraction}. From equation \eqref{kinematicEQ} we can obtain an expression for the two refraction angles:
%
%
\begin{equation}\label{refraction:angles}
\theta_\pm=\arcsin\left(\frac{k^\text{N}}{k_\pm^\text{SO}}\sin\phi\right)\,.
\end{equation}
%
%
If the refraction angle is equal to $\frac{\pi}{2}$, than the corresponding mode will not propagate  forward, thus it is \emph{closed}. Of course, there are two critical angles $\phi_\text{c}^\pm$ corresponding to the closure of the corresponding modes in the SO region defined by the relation:
%
%
\begin{equation}\label{critical:angles}
\phi_\text{c}^\pm=\arcsin\left(\frac{k_\pm^\text{SO}}{k^\text{N}}\right)\,.
\end{equation}
%
%
%
%
\begin{figure}
\centering
\includegraphics[width=\columnwidth]{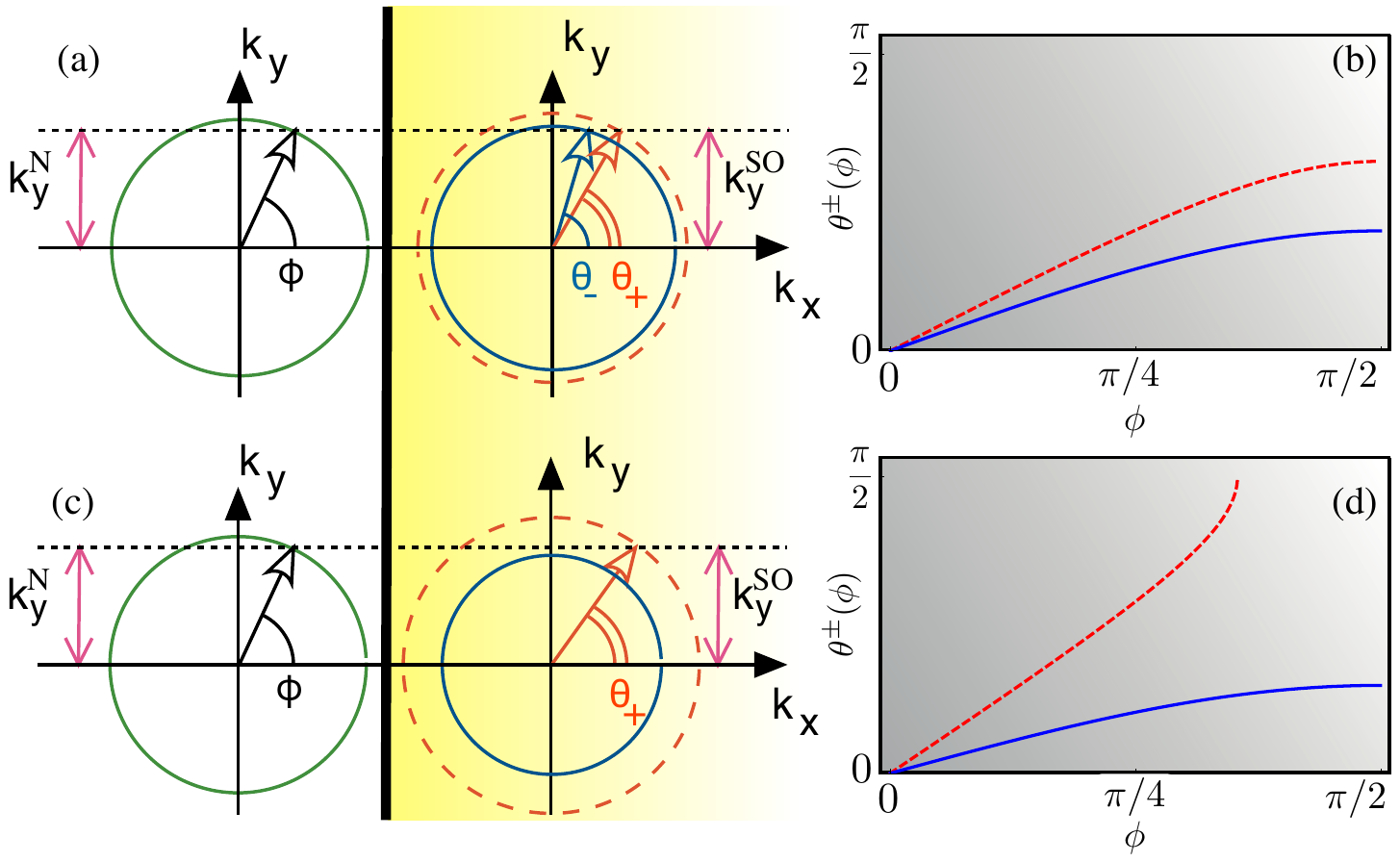}
\caption{\label{fig:angles}(Colours online) Spin Refraction angles $\theta_\pm$ as a function of the incoming particle angle $\phi$. Panel (a) and (b) case parameters of two modes always open. Panel (c) and (d) case parameters where the critical angle of mode (-), $\phi_\text{c}^-$ is bigger than $\pi/2$. In Panels (b) and (c) the red-dashed line is the mode (-) whereas the blue-solid line is the mode (+).}
\end{figure}
%
%
According to equation \eqref{kfermi}: $k_+^\text{SO}<k_-^\text{SO}$ so that $\phi_\text{c}^+<\phi_\text{c}^-$. Furthermore, considering that $k_+^\text{SO}$ is always smaller than $k^\text{N}$, the mode $(+)$ will be always open. The previous results are based only on kinematic considerations and are independent by the nature of the SOI. The key ingredient is the presence of an interaction that splits  the 2D Fermi surface in two circles. For example, the same physics would be possible at an interface between a normal region and a region with a strong Zeeman splitting, assuming that the Fermi energy allows the propagation along the two modes in the region with the magnetic field.

Some care is required when writing down the Hamiltonian for this scattering problem. Indeed the two regions (N and SO) are obtained by considering an inhomogeneous coupling constant for the RSOI $\alpha(x)$. As a consequence, $\alpha(x)$ does not commute  with the momentum operator $[\alpha(x),p_x]\neq0$. Thus the Hamiltonian describing the two regions has to read:
%
%
\begin{align}\label{ham:interface}
\mathcal{H}_\text{N-SO}=&\frac{\bm{p}^2}{2m^*} +\frac{\alpha(x)}{\hbar}(s_x p_y-s_y p_x)\nonumber\\
&-\rmi s_y \frac{1}{2m^*} \frac{\partial \alpha(x)}{\partial x} + V_\text{int}\delta(x).
\end{align}
%
%
where the second to last terms reestablish   the hermiticity~\cite{Messiah:2014}. In the simplest approximation we can consider $\alpha(x)=\alpha\Theta(x-x_\mathrm{int})$, where $\Theta(x)$ is the Heaviside step function. The last term mimics the presence an SOI interface, according to the standard approach describing interfaces with delta potentials of strength $V_\mathrm{int}$~\cite{blonder:1982}. A more realistic description has to account also for possible changes of the effective electron mass $m^*\to m^*(x)=m^*_\text{N}\Theta(x_\mathrm{int}-x)+m^*_\text{SO}\Theta(x_\mathrm{int}+x)$ in the two regions, more details can be found in Ref.~[\onlinecite{perroni:2007,marigliano:2003}]. When considering the presence of a second interface, where the RSOI vanishes again, the refraction angles for the two modes will be the same and equal to the incoming one $\phi$~\cite{khodas:2004,marigliano:2004}. However this double interface structure cannot produce a spin polarisation when  the incoming particles are unpolarised. This is due to the fact that in order to produce a spin polarisation we need to violate the Onsager relation for a two terminal system~\cite{Adagideli:2012,Gorini:2012}. Violation is possible if and only if time reversal symmetry is broken, this implies, \emph{e.g.}, the application of a magnetic field.

The same physics is possible also in the case of SLG, the main difference is that the RSOI in this case does not contain the momentum operator \eqref{soi:slg}. As a consequence we do not need the extra term in the full Hamiltonian as in equation \eqref{ham:interface}. A complete analysis of the spin-double refraction for the case of graphene is reported in Ref.~\cite{Bercioux:2010} where a { transfer matrix method is presented}, which is useful for studying the spin-dependent transport in hybrid structures in SLG, as a periodic RSOI potential~\cite{Lenz:2011,Costa:2013}. 

\section{Pure spin current generation in RSOI material}\label{spin:current} 

A very crucial point in spintronics is the creation of pure spin currents. Over the last 20 years, many methods for creating pure spin currents have been proposed. We will focus mainly on ratchet and pumping methods that have received also experimental verification. Here we  define what is a  spin current and  discuss a method for evaluating it in terms of the Landauer-B\"uttiker formalisms~\cite{Scheid:2007}. A pure spin current is defined as a particle flow carrying finite spin polarisation without an associated charge current.

A simple example follows: Suppose, for instance, that  a charge $\mathcal{Q}$ and spin polarisation $\mathcal{S}_{pq}$  moves from a contact $p$ to $q$ during the time $t\in [0,T]$ and in the next period  $t\in [T,2T]$  the same amount of charge $\mathcal{Q}$, but a different spin polarisation $\mathcal{S}_{qp}$,  moves from contact $q$ to $p$. The net charge transported between $p$ and $q$, in the time $t\in [0,2T]$ is zero, but the net spin polarisation is $\Delta\mathcal{S}=\mathcal{S}_{pq}-\mathcal{S}_{qp}$.

In order to evaluate the spin-current we consider $N$ non ferromagnetic contacts injecting spin-unpolarised current into the leads. We use, as customary, a local coordinate system for the lead under investigation, where $x$ is the coordinate along the lead in the direction of charge propagation due to an applied bias in linear response and $y$ is the transverse coordinate. In each lead, at a fixed energy $E$, several conducting modes are open. The wave function for each  mode reads
%
%
\begin{equation}
\Psi^{\pm} _{E,n s}(x,y)=\frac{1}{\sqrt{k_n(E)}}\text{e}^{\pm \text{i} k_n(E) x} \chi _n(y)\Sigma (s)\,,
\label{leadeigenfunct}
\end{equation}
%
%
here we have introduced the transverse eigenfunctions of the lead of width $W$:
%
%
\begin{equation*}
\chi _n(y)=\sqrt{\frac{2}{W}} \sin\left( \frac{n\pi y}{W}\right) \; \forall n\in\mathbb{N} 
\end{equation*}
%
%
with the eigenenergy $E_n=\frac{\hbar^2}{2m^*}\left(\frac{n \pi}{W}\right)^2$ and $\Sigma (s)$ is the spin eigenfunction. The superscript~$\pm$ of $\Psi$ refers to the motion direction along the lead axis with the wave-vector $k_n=\sqrt{2m^*(E-E_n)/\hbar^2}$. We use the scattering approach for deriving the current formula:  the amplitudes of the states inside the leads are related via the scattering matrix $\bm{S}(E)$, determined by the Hamiltonian of the coherent conductor. 
The scattering state inside the $q$-th lead reads 
%
\begin{equation*}
\varphi^q_E(x,y)= \sum_{n s }\left[a^{q}_{n s} (E)\Psi^+_{E,n s}(x,y) + b^{q}_{n s} (E)\Psi^-_{E,n s}(x,y)\right],
\end{equation*}
%
%
%
with $(s = \pm$). It consists of incoming states $\Psi^+$ entering the coherent conductor from the contact $q$ and 
outgoing states $\Psi^-$ leaving the coherent conductor into the contact $q$. 
The amplitudes of incoming $a^j_{n s}$ and outgoing $b^i_{n s}$ states are related each other via the scattering relation  
%
%
%
\begin{equation}
b^i_{n' s'}(E)=\sum^{N}_{j=1}\sum_{n \in j}\sum_{s=\pm 1} S^{i,j}_{n' s' ,n s}(E)a^j_{n s}(E),
\label{S_matrix_action_definition}
\end{equation}
%
%
%
where the scattering matrix $\bm{S}(E)$ has the following structure for an $N$ terminal system:
%
\begin{equation*}
\bm{S}(E)=
\left(
\begin{array}{cccc}
\bm{r}^{1,1}(E)&\bm{t}^{1,2}(E)&\cdots&\bm{t}^{1,N}(E)\\
\bm{t}^{2,1}(E)&\bm{r}^{2,2}(E)&\cdots&\bm{t}^{2,N}(E)\\
\vdots          &   \vdots        &\ddots&\vdots      \\
\bm{t}^{N,1}(E)&\bm{t}^{N,2}(E)&\cdots&\bm{r}^{N,N}(E)
\end{array}
\right).
\end{equation*}
%
%
%
Here the sub-matrix $\bm{r}^{j,j}(E)$ is a square matrix of dimensionality $M^j(E)$, corresponding to the number of open channels in lead $j$ at energy $E$ connected to a reservoir with chemical potential $\mu _j$ --- it already includes the spin degree of freedom. The matrix $\bm{r}^{j,j}(E)$ contains 
the scattering amplitudes of incoming channels of lead $j$ being reflected back into outgoing channels of the same lead. The sub-matrix $\bm{t}^{i,j}(E)$ is a $M^i(E)\times M^j(E)$ matrix that contains the scattering amplitudes for transmission between incoming channels from lead $j$ and outgoing channels of lead $i$.\\

In order to evaluate the spin-current we consider $N$ non ferromagnetic contacts injecting spin-unpolarised current into the leads. We use, as customary, a local coordinate system for the lead under investigation, where $x$ is the coordinate along the lead in the direction of charge propagation due to an applied bias in linear response and $y$ is the transverse coordinate. In each lead, at a fixed energy $E$, several conducting modes are open. The wave function for each  mode reads
%
%
\begin{equation}
\Psi^{\pm} _{E,n s}(x,y)=\frac{1}{\sqrt{k_n(E)}}\text{e}^{\pm \text{i} k_n(E) x} \chi _n(y)\Sigma (s)\,,
\label{leadeigenfunct}
\end{equation}
%
%
here we have introduced the transverse eigenfunctions of the lead of width $W$:
%
%
\begin{equation*}
\chi _n(y)=\sqrt{\frac{2}{W}} \sin\left( \frac{n\pi y}{W}\right) \; \forall n\in\mathbb{N} 
\end{equation*}
%
%
with the eigenenergy $E_n=\frac{\hbar^2}{2m^*}\left(\frac{n \pi}{W}\right)^2$ and $\Sigma (s)$ is the spin eigenfunction. The superscript~$\pm$ of $\Psi$ refers to the motion direction along the lead axis with the wave-vector $k_n=\sqrt{\frac{2m^*}{\hbar^2}(E-E_n)}$. We use the scattering approach for deriving the current formula:  the amplitudes of the states inside the leads are related via the scattering matrix $\bi{S}(E)$, determined by the Hamiltonian of the coherent conductor. 
The scattering state inside the $q$-th lead reads 
%
\begin{equation*}
\varphi^q_E(x,y)= \sum_{n s }\left[a^{q}_{n s} (E)\Psi^+_{E,n s}(x,y) + b^{q}_{n s} (E)\Psi^-_{E,n s}(x,y)\right],
\end{equation*}
%
%
%
with $(s = \pm$). It consists of incoming states $\Psi^+$ entering the coherent conductor from the contact $q$ and 
outgoing states $\Psi^-$ leaving the coherent conductor into the contact $q$. 
The amplitudes of incoming $a^j_{n s}$ and outgoing $b^i_{n s}$ states are related each other via the scattering relation  
%
%
%
\begin{equation}
b^i_{n' s'}(E)=\sum^{N}_{j=1}\sum_{n \in j}\sum_{s=\pm 1} S^{i,j}_{n' s' ,n s}(E)a^j_{n s}(E),
\label{S_matrix_action_definition}
\end{equation}
%
%
%
where the scattering matrix $\bi{S}(E)$ has the following structure for an $N$ terminal system:
%
\begin{equation*}
\bi{S}(E)=
\begin{pmatrix}
\bi{r}^{1,1}(E)&\bi{t}^{1,2}(E)&\cdots&\bi{t}^{1,N}(E)\\
\bi{t}^{2,1}(E)&\bi{r}^{2,2}(E)&\cdots&\bi{t}^{2,N}(E)\\
\vdots          &   \vdots        &\ddots&\vdots      \\
\bi{t}^{N,1}(E)&\bi{t}^{N,2}(E)&\cdots&\bi{r}^{N,N}(E)
\end{pmatrix}.
\end{equation*}
%
%
%
Here the sub-matrix $\bi{r}^{j,j}(E)$ is a square matrix of dimensionality $M^j(E)$, corresponding to the number of open channels in lead $j$ at energy $E$ connected to a reservoir with chemical potential $\mu _j$ --- it already includes the spin degree of freedom. The matrix $\bi{r}^{j,j}(E)$ contains 
the scattering amplitudes of incoming channels of lead $j$ being reflected back into outgoing channels of the same lead. The sub-matrix $\bi{t}^{i,j}(E)$ is a $M^i(E)\times M^j(E)$ matrix that contains the scattering amplitudes for transmission between incoming channels from lead $j$ and outgoing channels of lead $i$.\\
%
%
The wave function of the scattering state inside lead $i$, where only the incoming channel $(n s)\in j$ is populated 
$(a^{j'}_{n's'}=\delta _{j',j}\delta _{n',n}\delta _{s',s})$,  
reads for $j=i$:
%
%
\begin{align}
\varphi ^i_{E,n s}(x,y) = &
\Psi ^+_{E,n s}(x,y)\nonumber\\& +\sum_{(n's')\in i}r^{i,i}_{n's',n s} (E)
\Psi ^-_{E,n's'}(x,y),
\label{Mscattstatp}
\end{align}
%
%
and, correspondingly, for $j \neq i$
%
%
\begin{equation}
\varphi ^i_{E,n s}(x,y)=
\sum_{(n's')\in i}t^{i,j}_{n's',n s} (E) \Psi^-_{E,n's'}(x,y).
\label{Mscattstatq}
\end{equation}
%
%
For a generic spin wave function $\Phi(x,y)$ the spin current $I_{\mathrm{S}}(x)$ passing a cross 
section ($x=\,$const) of a lead is given by:
%
%
%
\begin{equation}
I_{\mathrm{S}}(x)=\int \mathrm{d}y \, \Phi^*(x,y)\bi{J}_\mathrm{S}\Phi(x,y).
\label{A:Eq:5}
\end{equation}
%
%
%
%
Here, we have used the most common definition of the spin current operator~\cite{Rashba:2006}. This is defined with respect to an arbitrary quantisation axis $\bi{u}$, than the operator $\bi{J}_{\mathrm{S}}$ reads:
%
%
\begin{equation}
\bi{J}_{\mathrm{S}}=\left(\frac{\hbar}{2}\bi{s}\cdot\bi{u}\right)\left[\frac{\hbar}{2m^*\rmi}\left( \left.\frac{\partial}{\partial x}\right|_{\rightarrow} - \left.\frac{\partial}{\partial x}\right|_{\leftarrow}\right)\right]\,. 
\end{equation}
%
%
In this operator we have the spin operator projected along the $\bi{u}$ quantisation direction times that standard quantum mechanical current. Here, as customary, the partial derivatives act on the expressions to their right and left  (as indicated by the arrows), respectively. By acting this the former operator on the scattering state \eqref{Mscattstatp} we obtain for the spin current~\eqref{A:Eq:5} inside lead $i_{n\sigma}$
%
%
\begin{equation*}
I^{j = i}_{\mathrm{S};E,n s}(x\in i)=\frac{\hbar ^2}{2m^*}\left[ s -
\sum_{(n's')\in i}s' \left|r^{i,i}_{n's',n s}(E)\right| ^2 
\right] \, ,
\end{equation*}
%
%
where $(n,s \in j, j = i)$. For the other scattering state \eqref{Mscattstatq} we find the corresponding expression ($n,s \in j,j \neq i$)
%
%
\begin{equation*}
I^{j\neq i}_{\mathrm{S};E,n s}(x\in i)=-\frac{\hbar ^2}{2m^*}\sum_{(n's')\in i}
s'\left|t^{i,j}_{n's',n s}(E)\right| ^2 \, .
\end{equation*}
%
%
Each channel is populated according to the Fermi-Dirac distribution $f(E;\mu _n)$ of the respective contact $n$, the statical average of the total spin current in lead $i$ reads
%
%
\begin{align}
\label{eq:I_S}
 \langle I_{\mathrm{S}}(x\in i)\rangle   = &  \frac{m^*}{2\pi\hbar ^2}\int_{0}^{\infty} \mathrm{d}E \; \nonumber \\ 
& \Bigg[ \sum_{j=1}^{N}\sum_{(n s)\in j}f(E;\mu _j ) I^{j}_{\mathrm{S};E,n s}(x\in i)
\Bigg] \\
& =  -\frac{1}{4\pi}\int_{0}^{\infty} \!\!\!\mathrm{d}E \Big[ f(E;\mu _i)R^{i,i}_{\mathrm{S}}(E) \nonumber \\
&+\sum_{q\neq i}f(E;\mu _q ) 
T^{i,q}_{\mathrm{S}}(E)\Big] \nonumber
\end{align}
%
where the spin resolved transmission and reflection are defined as
%
\begin{eqnarray*}
T^{i,q}_{\mathrm{S}}(E)& = & \sum_{s'=\pm}\left( T^{i,q}_{+,s'}-T^{i,q}_{-,s'}\right)\nonumber\\
R^{i,i}_{\mathrm{S}}(E)& = & \sum_{s'=\pm}\left( R^{i,i}_{+,s'}-R^{i,i}_{-,s'}\right)\,  ,\nonumber
\end{eqnarray*}
%
with
%
\begin{eqnarray}\label{TransDef}
T^{i,q}_{s,s'}(E) & = & \sum_{n\in i}\sum_{n'\in q}\left|t^{i,q}_{n s,n's'}(E)\right| ^2 \, ,\\
R^{i,i}_{s,s'}(E) & = & \sum_{n\in i}\sum_{n'\in i}\left|r^{i,i}_{n s,n's'}(E)\right| ^2 \, .
\end{eqnarray}
%
%
Probability conservation implies that the scattering matrix has to be unitary $[\bi{S}(E)]^\dag\bi{S}(E)=\bi{S}(E)[\bi{S}(E)]^\dag=\mathbb{I}$, than the following relation holds:
%
%
%
%
\begin{equation*}
\sum_{(n's')\in i}\!\!\left|r^{i,i}_{n s,n's'}(E)\right| ^2 +
\sum_{q\neq i}\sum_{(n''s'')\in q}\!\!\left|t^{i,q}_{n s,n''s''}(E)\right| ^2 = 1.
\end{equation*}
%
%
By using the symmetry relation between the spin transmission and reflection (c.f.~Ref.~[\onlinecite{Scheid:2007}]), it is straightforward to show that 
%
%
%
\begin{equation*}
R^{i,i}_{\mathrm{S}}(E)+\sum_{q\neq i}T^{i,q}_{\mathrm{S}}(E)=0\,.
\end{equation*}
%
In view of Eq.~(\ref{eq:I_S}) we eventually find for the spin current in lead $i$
%
%
\begin{equation}
I_{\mathrm{S}}(x\in\! i)=\!
\frac{1}{4\pi}\int_{0}^{\infty}\!\! \mathrm{d}E 
\sum_{q\neq i}  
\left[ f(E;\mu _i)\!-\! f(E;\mu _q ) \right]
T^{i,q}_{\mathrm{S}}(E) .
\label{spincurr}
\end{equation}
%
%
Equilibrium spin currents can locally exist in systems with SOI as shown for 2DEGs~\cite{Rashba:2003} and in mesoscopic systems~\cite{Nikolic:2006}, however
equation \eqref{spincurr} clearly shows that at thermal equilibrium ($\mu _j=\mu~~\forall j\in\{1,N\}$) the spin current vanishes inside leads without SOI and/or magnetic field. This absence of equilibrium spin currents in the leads has been shown for systems with preserved time-reversal symmetry~\cite{Kiselev:2005}. 
An expression very similar to \eqref{spincurr} has been used to investigate the  SHE~\cite{Hankiewicz:2004,Ren:2006,Bardarson:2007,Brune:2012}. In the presence of SOI or magnetic fields it is nontrivial to write down a continuity equation for the spin current, the problem has been addressed in several articles~\cite{Nikolic:2006,Sun:2005,Shi:2006,An:2012}.
%
%
%
\subsection{Pure spin current generators}

In this section we  describe two  mechanisms that can be used for generating a pure spin current.

\subsubsection{Spin Ratchet}

A particle ratchet~\cite{Reimann:2002,Haenggi:2009} is  a system with broken inversion (left/right) symmetry that generates a net currents upon external \emph{ac}-driving in the absence of a net (time-averaged) bias potential. Ratchets  have much in common with current rectifiers, though there are differences, in particular
in the dissipative case~\cite{Reimann:2002}. The theoretical concept of ratchets, originally introduced for classical dynamics, was later extended to the quantum dissipative regime~\cite{reimann:1997}. The main difference  between ratchets and rectifiers is that  quantum ratchets exhibit current reversal upon changing, \emph{e.g.}, the temperature or energy~\cite{reimann:1997,Haenggi:2009}.
Such quantum ratchets were experimentally realised in semiconductor heterostructures  in a chain of asymmetric ballistic electron  cavities in the low-temperature regime, with a dynamics was close to coherent~\cite{Linke:1999}. Also the charge current reversal phenomenon has been demonstrated~\cite{Linke:2002}.

A spin ratchet is a quantum ratchet with \emph{lack} of broken spatial symmetry --- in order to get a zero charge current --- and a finite RSOI ensuring a breaking of the spin symmetry thus allowing for a finite spin current. Theory predicts spin ratchets to work both in the ballistic regime~\cite{Scheid:2007b,Scheid:2010,Ang:2015} and in the dissipative one~\cite{Smirnov:2008a,Smirnov:2008b,Smirnov:2009}. A set-up for the ballistic ratchet has been also proposed by using periodically disposed  magnetic stripes instead of electrostatic barriers and RSOI~\cite{Scheid:2007,Scheid:2006}.

\begin{figure}[!ht]
\centering
\includegraphics[width=\columnwidth]{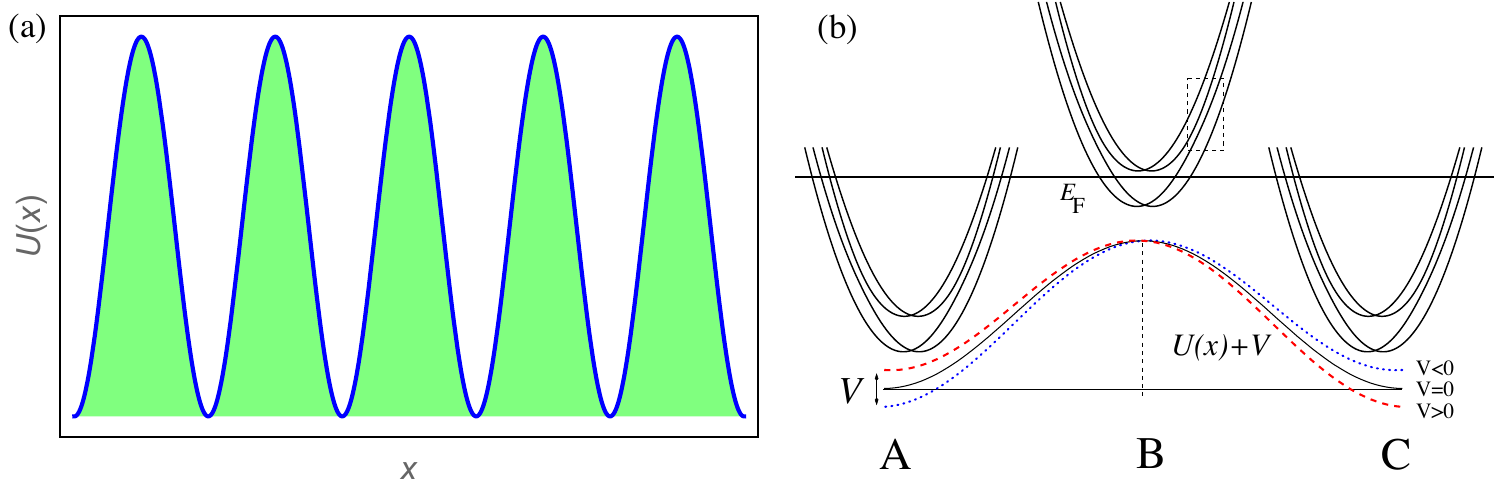}
\caption{\label{spin:ratchet}(Colours online)  Panel (a): Sketch of the system of symmetric periodic equally spaced barriers used for the ballistic spin ratchet.
Panel (b): Illustration of the spin polarisation mechanism for transmission through a strip with a single adiabatic symmetric potential barrier $U(x)$ (solid line) in the two rocking situations (dashed and dotted line). At points A, B and C the position-dependent energy dispersion relation $E_n(k_x)$ is sketched with respect to the Fermi energy $E_\text{F}$ (horizontal line) for two transverse modes and RSOI-induced spin splitting of each mode. Courtesy of Ref.~[\onlinecite{Scheid:2007b}]}
\end{figure}

A ballistic spin ratchet is mainly constituted by a quasi-one dimensional channel with a set of symmetric barriers and RSOI [see \fref{spin:ratchet}(a)]. The lack of broken symmetry  implies the absence of a net charge current. In order to understand the presence of a finite spin current we show here the same argument proposed in Refs.~[\onlinecite{Scheid:2007b,Scheid:2010}]. A key ingredient is the sub-band hybridisation that we have introduced in Sec.~\ref{RSOI2DEG}.

We consider a wire with two open transverse modes $(N=1,2)$ and a smooth symmetric potential barrier $U(x)$ in the two bias or \emph{rocking} situations ($V\gtrless0$), see \fref{spin:ratchet}(b).
%
%
%
%
Upon adiabatically traversing the barrier from region A via B to C [see lower part of \fref{spin:ratchet}(b)], the electron energy spectrum $E_n(k_x)$ split by RSOI is shifted rigidly upward and downward by the adiabatic potential barrier~$U(x)$. For fixed Fermi energy  $E_\text{F}$, the initial shift causes a depopulation of the upper levels ($N=2$) and a spin-dependent repopulation of the lower one ($N=1$) while moving from B to C. If $E_\text{F}$ is traversing an anti-crossing between successive modes (see the region indicated by the dashed window in \fref{spin:ratchet}(b), there is a certain probability $P$ for  electrons to change their spin state. This causes an asymmetry between spin-up and spin-down states for the repopulated levels~\cite{EtoMikio:2013}. The related transition probability can be computed in a Landau-Zener  framework. In the case of a transverse parabolic confinement of frequency $\omega_0$, this probability reads~\cite{EtoMikio:2013,Scheid:2007b}:
%
%
\begin{equation}
P(\pm V_0)\!=\! 1-\exp \left\{\frac{-\pi k_\text{SO} \omega_0\Sigma_z^{-1}}{
\partial_x [U(x,z)\pm V_0 g(x,z)]} \right\} \, .
\label{landau-zener}
\end{equation}
%
%
Here  $\Sigma_z$ denotes the difference in the polarisation of the two modes involved,  and the function $g(x,z)$  describes how the potential drops inside the system. In principle $g(x,z)$ is obtained by solving the Boltzmann equation, however its exact form is irrelevant in order to understand the working principle of the spin ratchet. The spin transmission is proportional to $P(V)$ and thus different for the two rocking situations $V\gtrless0$. Hence, the ratchet spin current $\langle I_\text{S} (V)\rangle$ is finite, also for the case of a symmetric barrier.  A quantitative explanation of the spin ratchet effect for a system of multiple barriers [\fref{spin:ratchet}(a)] is beyond this model. An experimental verification of the spin ratchet effect was proposed in Ref.~[\onlinecite{Costache:2010}]. However, in this experiment the breaking of the spin symmetry was achieved by combining a superconductor with a magnetic field. 

\subsubsection{Spin Pumping}

Adiabatic charge pumping~\cite{Brouwer:1998} consists of the transport of charge obtained --- at zero bias voltage --- through the periodic modulation of some parameters in the scattering region, \emph{e.g.} some voltages. 

In 2003, Governale \emph{et al.}~[\onlinecite{Governale:2003}] proposed an updated of the scheme of charge pumping extended to the spin, following a suggestion by Mucciolo \emph{et al.}~[\onlinecite{Mucciolo:2002}]. The pumping scheme consists of an electrostatic barrier that is changed periodically in time and a gate,  as second pumping parameter, to tune the strength of RSOI~\cite{Governale:2003}. They showed that, neglecting sub-band hybridization, the charge current is zero whereas the spin current is finite. However, in a more realistic case in which many modes are opened, and consequently coupled by RSOI, the charge current is finite but two order of magnitude smaller than the spin one. A prototypal spin pumping system has been realised in a GaAs quantum dot in Ref.~[\onlinecite{Watson:2003}], in which an ingenious method for detecting spin currents is presented.
Similar pumping mechanisms have been proposed for graphene with RSOI~\cite{Bercioux:2012}, and TIs~\cite{Citro:2011,Inhofer:2013,Hofer:2013,Hofer:2014}.

\subsection{Datta and Das spin Field Effect Transistor}\label{SPINFILTER}
%
%
\begin{figure}
	\centering
	\includegraphics[width=\columnwidth]{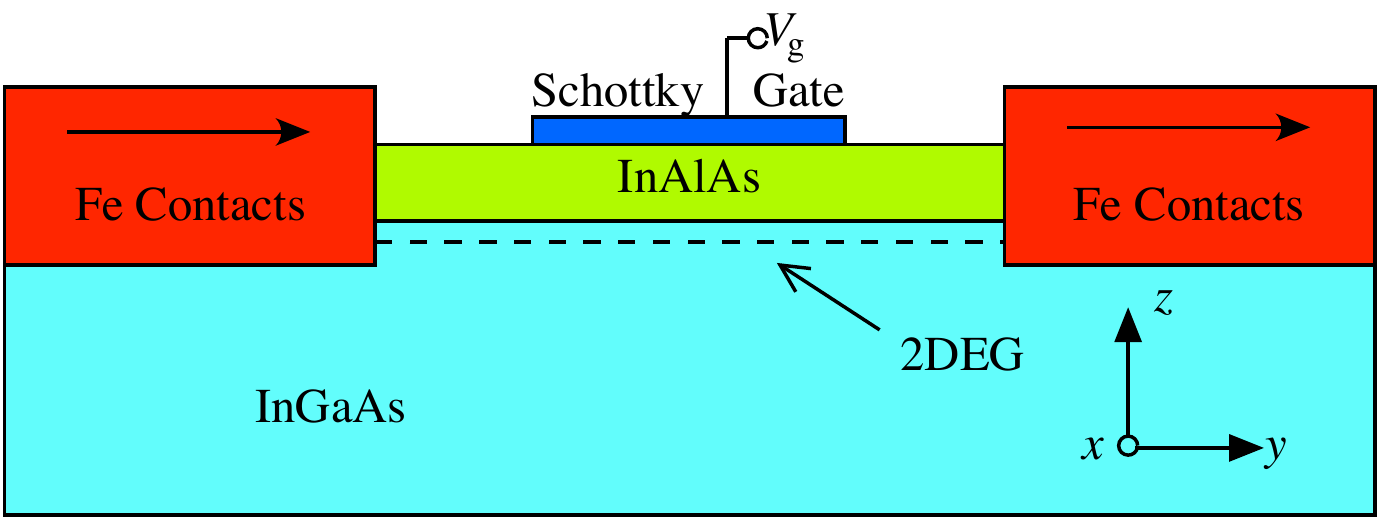}
	\caption[The spin-field effect transistor proposed by Datta
	and Das.]{(Colours online) The spin-field effect transistor proposed by Datta
	and Das~[\onlinecite{datta:1990}].\label{SFET-dd}}
\end{figure}
%
%
The spin momentum locking due to the RSOI is one of the key ingredients used in 1990 for the proposal of a spin-FET by Datta and Das~[\onlinecite{datta:1990}].   It is believed that spin-FET has the advantages of low energy consumption and fast switching speed since it does not involve creating or eliminating the electrical conducting channels during the switching, required by traditional FETs. This spin-FET is the electronic analog of an electro-optic modulator. The electro-optical effect makes the dielectric constant of a medium different along the two perpendicular directions (\emph{e.g.}, $y,z$).

Let us assume to have photons with a polarisation at 45$^\mathrm{o}$ with respect to the $y$ axis (in the $y$--$z$ plane)~\cite{Born:1999}. They can can be represented as a linear combination of $z$- and $y$-polarised photons:
%
%
\begin{equation}
\left(
\begin{array}{c}
1\\
1
\end{array}\right) _{45^\mathrm{o}}= 
\left(
\begin{array}{c}
1\\
0
\end{array}\right)_z +\left(
\begin{array}{c}
0\\
1
\end{array}\right) _y\,.
\end{equation}
%
%
Because of the anisotropy of the dielectric constant, as the light passes through the electro-optic material of length $L$, the two
components acquire different phase shifts $k_1 L$ and $k_2 L$.
Thus the light emerging from the electro-optic material can be represented as  $\left(
\mathrm{e}^{\mathrm{i} k_1 L},
\mathrm{e}^{\mathrm{i} k_2 L}
\right)^\mathrm{T} $. It is collected by an analyser polarised along the 
45$^\mathrm{o}$  direction. 
Therefore the output power is  given by  
%
%
\begin{equation}\label{powerDD}
P_0 \propto \left| (1~1) 
\left(\begin{array}{c}
\mathrm{e}^{\mathrm{i} k_1 L} \\
\mathrm{e}^{\mathrm{i} k_2 L}
\end{array}
\right)
\right|^2 = 
4 \cos^2 \displaystyle\frac{(k_1-k_2)L}{2} .
\end{equation}
%
%
The light output is modulated with a gate voltage that controls the
differential phase shift $\Delta \theta = (k_1-k_2)L$.

The analog electronic device based on the RSOI is shown in \fref{SFET-dd}.  In the original proposal by Datta and Das, the polariser and the analyser 
were  suggested to be implemented via ferromagnetic contacts (as Fe)~\cite{johnson:1988}. In such materials the density of states for electrons with a specific spin orientation --- at  the Fermi energy --- greatly exceeds that for the opposite direction. A contact magnetised in the $x$ direction preferentially injects and detects electrons spin polarised along positive $x$ which is represented as a linear combination of the positive $z$-polarised $|\!\uparrow\rangle$ and negative $z$-polarised electrons $|\!\downarrow\rangle$:
%
%
\begin{equation}
\left(
\begin{array}{c}
1\\
1
\end{array}\right) _{x}= 
\left(
\begin{array}{c}
1\\
0
\end{array}\right)_z +\left(
\begin{array}{c}
0\\
1
\end{array}\right) _z
\end{equation}
%
%

Finally, the analogue of the electro-optic material is realised by employing a 1D quantum wire with RSOI. Within a single band model, RSOI causes the $|\!\uparrow\rangle$ and $|\!\downarrow\rangle$ electrons with the same energy to have different wave vectors $k_\uparrow$ and $k_\downarrow$ [see \fref{spectrum2d}(e)]. Let us consider an electron travelling in the $x$ direction with $k_z=0$ and $k_x\neq 0$. The eigenenergies for the corresponding 1D case are given equation \ref{spectrumRashba} with $k_y=0$.

Using equations \eqref{spectrumRashba}-\eqref{kfermi}, it is possible to recover a relation for the phase shift between the two spins:
%
%
\begin{equation}\label{phase:shift}
\Delta \theta = \left( k_\uparrow -k_\downarrow \right) L=
\displaystyle\frac{2m \alpha L}{\hbar^2}
\end{equation}
%
%
which is proportional to $\alpha$.

The above analysis was originally limited to a single-mode quantum wire. However, as we have learnt in the Sec.~\ref{sub:Hyb}, in the presence of multiple modes, physics changes.

A fully numerical multi-mode analysis of the spin-FET  was proposed by Mireles and Kirczenow~\cite{mireles-2001}.  They investigate the effect of the strength of the RSOI on the spin-transport properties of narrow quantum wires of width $W$. 
%
%
\begin{figure}[!t]
	\centering
	\includegraphics[width=0.65\columnwidth]{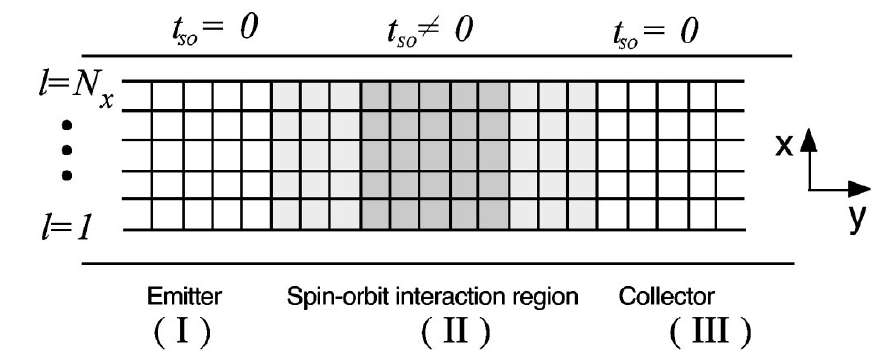}
	\caption[Schematic of the tight-binding model for the
	system.]{Schematic of the tight-binding model for the
	system. In the shaded areas the spin-orbit interaction if
	finite $t_{\mathrm{SO}} \neq
	0$. Courtesy of Ref.~[\onlinecite{mireles-2001}]. \label{mireles-1}}
\end{figure}
%
%
This is a quasi-one-dimensional wire, {\emph{i.e.}}, assumed to be infinitely long in the propagation direction (see \fref{mireles-1}). 

The nearest-neighbour tight-binding Hamiltonian the RSOI \eqref{Ham:RSOI} takes the form
\begin{widetext}
%
%
\begin{align}\label{rashba-tb}
\mathcal{H}_{\mathrm{SO}}^{\mathrm{tb}}(y)  =  -
\displaystyle t_\mathrm{SO}\sum_{s,s'} \sum_{\langle \ell,\ell'\rangle_y}\sum_{\langle j,j'\rangle_x} \left(
c_{\ell,m,s'}^\dag \left(\mathrm{i} s_y\right)_{s,s'}
c_{\ell',m,s}  - c_{\ell,j,s'}^\dag
\left(\mathrm{i} s_x \right)_{s,s'} c_{\ell',j',s} \right)
\end{align} 
%
%
\end{widetext}
where the summation is intended over next-neighbours $\langle \ldots\rangle_{x,y}$, $t_{\mathrm{SO}} = \frac{\alpha}{2} a$ is the isotropic next-neighbour transfer RSOI hopping term  (where $a$ is the lattice constant), and $c_{\ell,j,s}^\dag$ the electron creation operator on the site ($\ell,j$) with spin state $s$ ($s=\uparrow,\downarrow$).

The wire is divided in three main regions. In two of these [I and III in \fref{mireles-1}], which are near the ferromagnetic source and drain, the  parameter $t_{\mathrm{SO}}$ is set to zero. In the middle region (II) the RSOI is finite ($t_{\mathrm{SO}}\neq 0$). 

The spin-dependent transport problem is solved numerically through the use of the spin-dependent Lippman-Schwinger equation\footnote{Nowadays a lot of numerical computation, also for systems with RSOI, is performed via recursive Green's function method; among the free codes one of the most used is KWANT~\cite{Groth:2014}.}
%
%
\begin{equation}
| \Psi \rangle = | \Phi \rangle + G_0(E)
\hat{\mathcal{H}}_{\mathrm{SO}}^{\mathrm{tb}} | \Psi \rangle, 
\end{equation}
%
%
where $| \Phi \rangle$ is the unperturbed wave function, \textit{i.e.} an eigenstate of the kinetic lattice Hamiltonian $\mathcal{H}_0$ without RSOI, and $G_0(E) = (E+\text{i} \epsilon - \mathcal{H}_0)^{-1}$ is the Green's functions for the system in the absence of any kind of scattering.

A  simple criterion for distinguishing the cases of \textit{weak} and \textit{strong} RSOI is introduced in Ref.~\cite{mireles-2001}. The contribution of the mixing of the spin sub-bands should be negligible as long the sub-bands spacing $\Delta E_\mathrm{W} = E_m^0 -E_n^0$ is much larger than the sub-band intermixing energy
%
%
\begin{equation}\label{energy-mixing}
\frac{\langle \phi_{n,s} | \mathcal{H}_{\mathrm{SOI}} |
\phi_{m,s'}\rangle}{E_m^0-E_n^0} \ll 1
\end{equation}
%
%
where $|\phi_{n,s}\rangle$ are the unperturbed electron wavefunctions. However, if the confinement energy and/or the RSOI are of the same order as the energy shift introduced by the sub-band mixing contribution, then the above condition is about one or larger. In this case one can introduce 
%
%
\begin{equation}\label{beta}
\beta_{\mathrm{SO}} \approx  \frac{\left(\frac{\pi a}{W}\right)^2}{\left(
\frac{\pi a}{W}\right) + a k_\mathrm{F}} = \beta_{\mathrm{SO}}^{\mathrm{c}},
\end{equation}
%
%
%
%
\begin{figure}[!t]
	\centering
	\includegraphics[width=0.8\columnwidth]{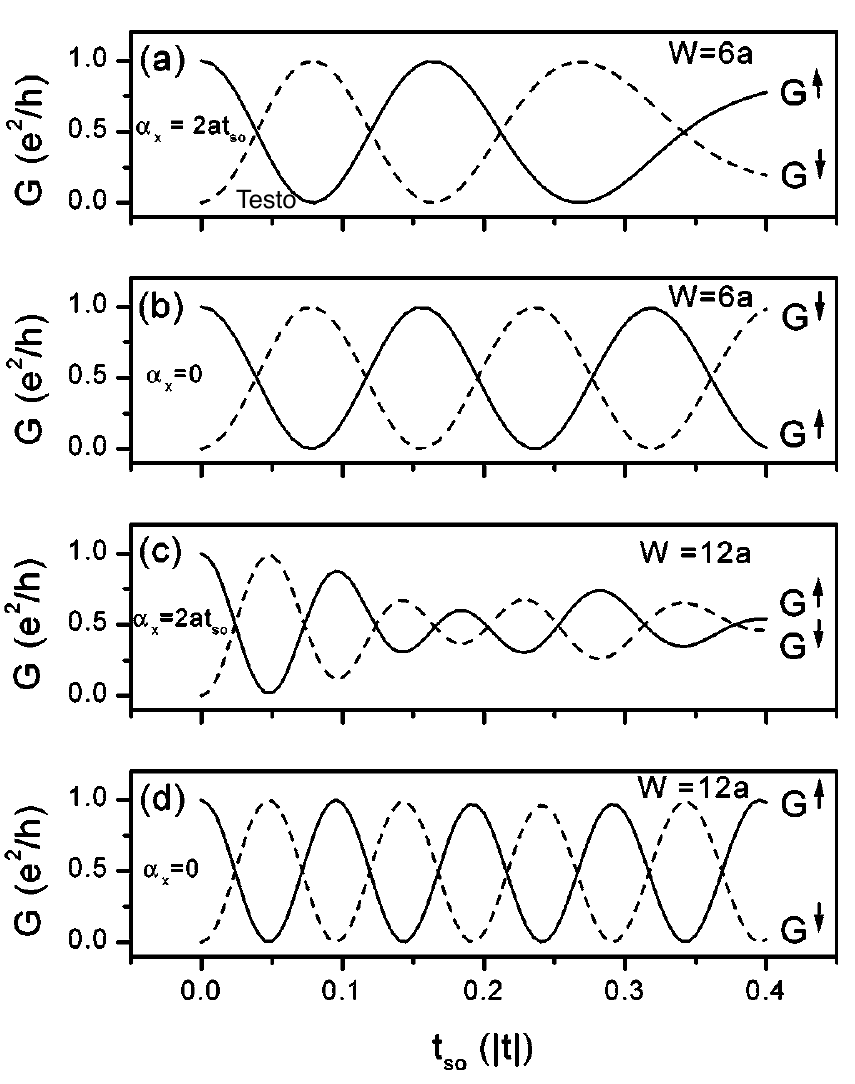}
	\caption[Spin-orbit coupling strength dependence of the ballistic spin conductance.]{Spin-orbit coupling strength dependence of the ballistic spin conductance; solid line is	$G^\uparrow$, dashed line $G^\downarrow$: (a) Narrow wire of 	$W=6a$ and uniform RSOI ($\alpha_x=\alpha_y=2 a
	t_{\mathrm{SO}}$). (b) Same as in (a) but with $\alpha_x=0$ and $\alpha_y=2 a t_{\mathrm{SO}}$; perfect oscillations are seen for all $t_{\mathrm{SO}}$. (c) Same as in (a) with $W=12a$. (d) Modulation for $W=12a$, with $\alpha_x=0$ and $\alpha_y=2 a t_{\mathrm{SO}}$. The sub-band mixing clearly changes the otherwise perfectly sinusoidal spin-conductance modulation. Courtesy of Ref.~[\onlinecite{mireles-2001}]. \label{mireles-2}}
\end{figure}
%
%
where $\beta_{\mathrm{SO}}= \frac{t_{\mathrm{SO}}}{|t|}$, and $k_\mathrm{F}$ is the Fermi wave number. The critical value $\beta_{\mathrm{SO}}^{\mathrm{c}}$ defines a \textit{weak} RSOI regime if $\beta_{\mathrm{SO}}<\beta_{\mathrm{SO}}^{\mathrm{c}}$ and a \textit{strong} coupling regime if $\beta_{\mathrm{SO}}>\beta_{\mathrm{SO}}^{\mathrm{c}}$~\cite{mireles-2001}. 

In \fref{mireles-2} the behaviour of the spin-resolved conductance as function of the RSOI parameter $t_{\mathrm{SO}}$  is shown. In \fref{mireles-2}(a) the incident Fermi energy is fixed to 0.5 ($k\approx 0.7 a^{-1}$) and $W=6a=60$~nm, which gives a critical value $\beta_{\mathrm{SO}}^{\mathrm{c}} = 0.22$. This value of $\beta_{\mathrm{SO}}$ separates the sinusoidal behaviour of $G^{\uparrow\downarrow}$ for $\beta_{\mathrm{SO}} \le 0.22$ from its behaviour for $\beta_{\mathrm{SO}} >  0.22$ where the confinement energy is of the order of the sub-band mixing energy. The effect is clearer for a wider wire ($W=120$~nm) [see \fref{mireles-2}(c)] for which the critical value of $\beta_{\mathrm{SO}}$ is $0.07$. To show that the non-sinusoidal behaviour is due mainly to the sub-band mixing, in the \fref{mireles-2}(b) and \ref{mireles-2}(d) we report the spin-conductance as function of $t_{\mathrm{SO}}$ with the same parameter of \fref{mireles-2}(a) and \ref{mireles-2}(c) respectively but in the unphysical situation of $\alpha_y\neq 0$ and $\alpha_x=0$, respectively. It is evident that the sinusoidal behaviour is recovered. {The modification of spin-FET due to the inclusions of many transversal modes has been also investigated in Ref.~[\onlinecite{Jeong:2006}].}

So far, several obstacles have been found on the way of the realisation of the spin-FET~\cite{datta:1990}. The main one is related to the injection of spin polarised currents. For example, it has been shown that in diffusive transport regime, for typical ferromagnets only a current with a small polarisation can be injected into a semiconductor 2DEG with long spin-flip length even if the conductivity of semiconductor and ferromagnet are equal~\cite{schmidt:2000}. This situation is dramatically exacerbated when ferromagnetic metals are used; in this case the spin polarisation in the semiconductor is negligible.

A possible solution to this problem was proposed by employing dilute magnetic semiconductor~\cite{ohno:1998} as source and drain contacts. In these systems a few percent of the cations in the III-V or II-VI semiconductors compounds are randomly substituted  by magnetic ions, usually Mn, which have local magnetic moments. The effective coupling between these local moments is mediated by free carriers in the host semiconductor compound (holes for $p$-doped
materials and electrons for $n$-doped one) and can lead to ferromagnetic long-range order. Curie temperatures $T_\mathrm{c}$ close to 100 K have been found in bulk (Ga,Mn)As systems~\cite{ohno:1998}.

Using the properties of the dilute magnetic semiconductor, all-semiconductor spin-FET have been proposed in which the conducting channel is provided by a two-dimensional hole gas~\cite{pala:2004}.

%
%
\begin{figure}[!t]
\begin{center}
\includegraphics[width=\columnwidth]{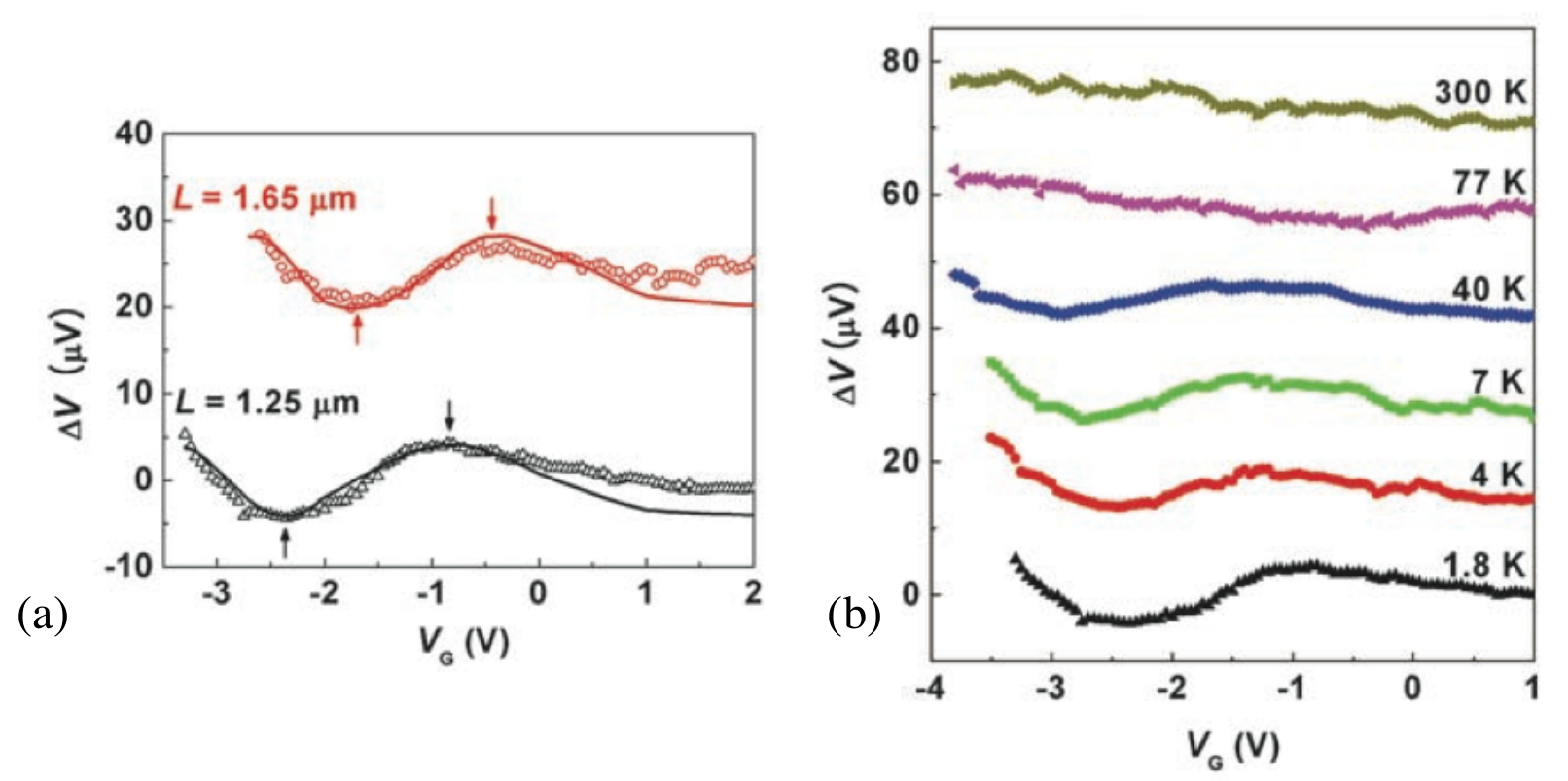}
\caption{\label{exp-DD}(Colours online) (a) Gate voltage modulation of spin FETs having different channel lengths, with T = 1.8 K and I = 1 mA. The symbols indicate experimental data. The solid lines are the fits obtained from $V=A \cos(2m^*\alpha L\hbar^{-1}+\varphi)$. Data are offset for clarity. Baseline voltages are 1.032 mV and 0.715 mV for $L$ = 1.25 $\mu$m and 1.65 $\mu$m, respectively; (b) Temperature dependence of oscillatory conductance with $L$ = 1.25 $\mu$m and $I$ = 1 mA. As temperature increases, the mean free path decreases and transport characteristics change from ballistic to diffusive. Courtesy of Ref.~[\onlinecite{koo:2009}].}
\end{center}
\end{figure}
%
%
The first experimental realisation of the spin-FET has been done in 2009 by the Johnson's group~\cite{koo:2009}. Instead of using the configuration initially suggested by Datta and Das~\cite{datta:1990}, in this experiment a nonlocal measurement scheme has been used~\cite{johnson:1988,Jedema:2001,Jedema:2002,Lou:2007}. This allows to perform a better measurement of the charge and spin signals. The quantum wire was realised in a InAs heterostructure with strong RSOI with the ferromagnetic contacts realised with Ni$_{81}$Fe$_{19}$ permalloy on the top.  \Fref{exp-DD}(a)  shows the oscillation of the output voltage of the spin-FET as a function of the gate voltage that is changing the strength of the RSOI for two different length of the distance between the two ferromagnetic contacts. As reported in equation \eqref{phase:shift} the phase shift between the two spin channels is proportional to the distance between the two ferromagnets. Therefore, for the longer case (red-solid data) we observe a shift of the length on the half period of oscillations.  \Fref{exp-DD}(b)  shows the oscillation for a fixed distance between the ferromagnetic contacts but at different temperatures. We observe a signature of coherent oscillations up to 40 K. At higher temperatures inelastic scattering become more pronounced and coherent effects are washed out~\cite{koo:2009}.  Another  experimental realization was also lately reported in Ref.~[\onlinecite{Chuang:2014}].
 
\section{Interference effects and Berry phase}\label{berry}

\subsection{Rashba interaction as SU(2) gauge field: application to quantum networks}
In this section we  explore RSOI  as a SU(2)  non-Abelian gauge field. We start by recasting the RSOI in the Hamiltonian~\eqref{rashba2DEG} as a SU(2) vector potential: 
%
%
\begin{eqnarray}\label{Ham:AC}
\mathcal{H}_0 & = & \frac{1}{2m} \left[ \bi{p} + \frac{m^* \alpha}{\hbar} (\bi{z}\times\bi{s})\right]^2 -\left(\frac{m^* \alpha}{\hbar}\right)^2 \nonumber\\
			& = &  \frac{1}{2m^*} \left( \bi{p} +  \mathcal{A}_{\bi{s}} \right)^2 -\left(\frac{m^*\alpha}{\hbar}\right)^2
\end{eqnarray}
%
%
comparing the two lines of \eqref{Ham:AC} we recognise $\mathcal{A}_{\bm{s}}=\frac{m^* \alpha}{\hbar} (\bm{z}\times\bm{s})$.  In this form the Rashba Hamiltonian  resembles that of a particle in a magnetic field, thus allowing for a straightforward connection with the physics of the  AB effect~\cite{aharonov:1959}. In the case of a SU(2) vector potential, it is known as AC effect~\cite{aharonov:1984}. In analogy to AB effect, we can introduce a phase field that reads:
%
%
\begin{equation}
	\psi_{\text{AC}} \equiv \frac{2\pi}{\phi_0} \oint \mathcal{A}_{\bm{s}}\cdot
	d\bm{r} = 2\pi\frac{\phi_{\text{SOI}}}{\phi_0},
\end{equation}
%
%
where $\phi_\text{SOI}$ is the flux associated to the RSOI effective field and $\phi_0=\frac{hc}{e}$ the flux quantum. When travelling through a closed path an electron gains a non-Abelian phase due to the presence of the RSOI. This extra phase can give rise to interference phenomena. By contrast to the phase gained in a perpendicular magnetic field --- that depends only on the area enclosed by the particle path--- here 
the actual path covered by the electron plays an important role. In the standard AB set-up the magnetic field can be tuned in order to move from a full destructive to a full constructive interference. In the case of AC the  role of the magnetic field is played by the RSOI that can be modulated  modifying $\alpha$, \emph{i.e.} by gating the heterostructure.  
The AC effect has been observed in semiconductor heterostructure~\cite{Bergsten:2006}, in HgTe rings~\cite{Konig:2006} and TI interferometric structures~\cite{Qu:2011} (c.f. \fref{Bi2Se3}). The interplay between the AB and the AC effect, and the mutual effect of an Abelin and non-Abelin gauge field has been investigated in Ref.~[\onlinecite{Nagasawa:2012,Nagasawas:2013}]. 

In order to understand the fundamental difference between the AB and the AC effect, we propose here a very simple \emph{gedankenexperiment} for a square interferometer. Let $\mathcal{R}_{pq}$ be the phase gained by the wave function $\Psi(\bm{r})$ --- travelling from a point $p$ to a point  $q$. It reads
%
%
\begin{subequations}\label{phases}
\begin{equation}\label{phaseSU2}
\mathcal{R}_{pq}^\text{SU(2)}=\exp\left\{ -\rmi \int_p^q\bm{s}\cdot(\bm{z}\times d\bm{l}) k_\text{SO}) \right\} 
\end{equation}
\begin{equation}\label{phaseU1}
\mathcal{R}_{pq}^\text{U(1)}=\exp\left\{ -\rmi \frac{2\pi}{\phi_0}\int_p^q \mathcal{A}\cdot d\bm{l}  \right\}
\end{equation}
\end{subequations}
%
%
depending on whether we are considering  a U$(1)$ or SU$(2)$ gauge field, {where $\mathcal{A}$ is the vector potential associated to a magnetic field $\bm{B}$}. 
%
%
\begin{figure}[!tb]
\begin{center}
\includegraphics[width=\columnwidth]{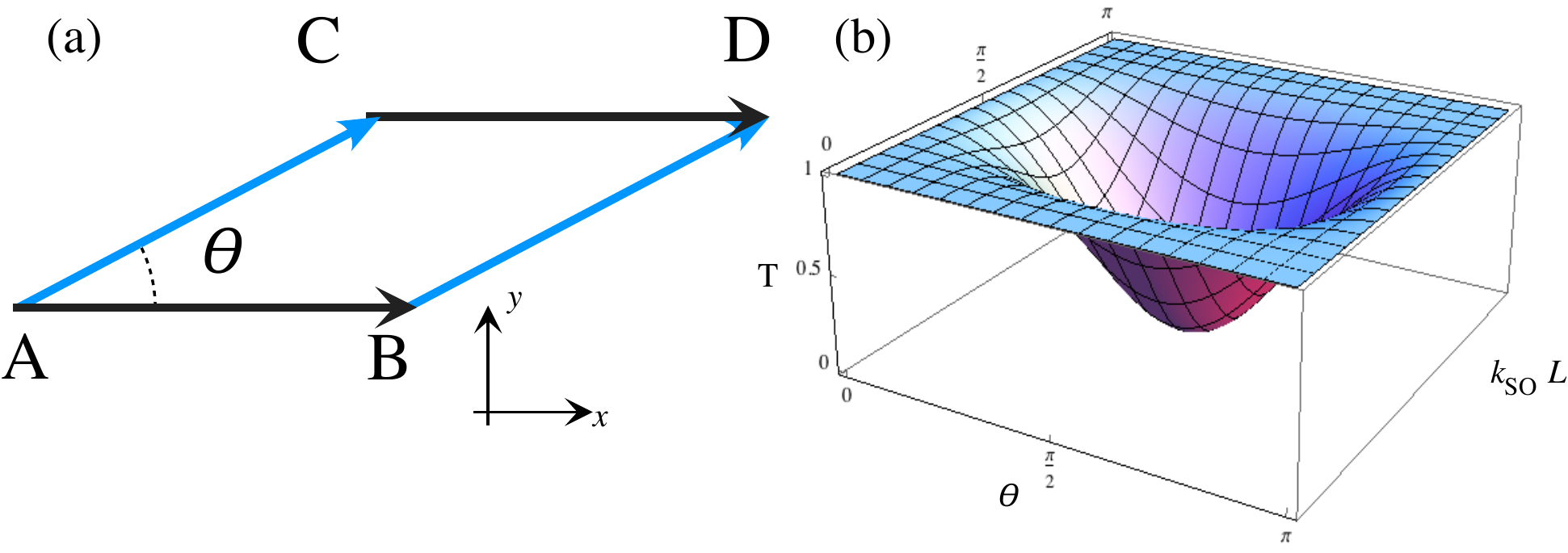}
\caption{\label{interferometer}(Colours online) (a): Scheme of the interferometer for addressing the difference between an Abelian and a non-Abelian phase. (b): Normalised transmission probability $T\propto\Tr[\Gamma\Gamma^\dag]$ for the interferometer in the left panel in the case of the action of a SU(2) gauge field \eqref{phaseSU2}.}
\end{center}
\end{figure}
%
%
We now consider the closed path in  \fref{interferometer}(a) and explicitly consider the phases  gained by the electron  travelling along each of its arm. The four arms of the interferometer have the same length $\ell_{pq}=L$  ($pq\in\{\text{AB,AC,BD,CD}\}$), the angle at its base is $\theta$, and its area is $S$. In order to obtain destructive interference the sum of the phases gained along consecutive arms has to be zero:
%
%
\begin{equation}\label{ABC:condition}
\Gamma=\mathcal{R}_\text{BD}^\beta\cdot\mathcal{R}_\text{AB}^\beta+\mathcal{R}_\text{CD}^\beta\cdot\mathcal{R}_\text{AC}^\beta=0
\end{equation}
%
%
with $\beta\in\{\text{SU(2),U(1)}\}$. 

By using the Gauss' theorem and defining the magnetic field flux as $\phi=|\bm{B}| S=(\nabla\times \mathcal{A}) S$, we obtain that in the case of a U(1) phase  the destructive interference is verified when the condition $\frac{\phi}{\phi_0}=\frac{1}{2}$ is met for every the value of the angle $\theta$. Here, the condition for the destructive interference is straightforward because the phases \eqref{phaseU1} are c-numbers.

In the case of the non-Abelian phase field the phases \eqref{phaseSU2} are linear combination of Pauli matrices. Therefore a solution for the condition \eqref{ABC:condition}  {is} obtained by looking at the transmission probability --- proportional  (to lowest order) to $\Tr[\Gamma\Gamma^\dag]$. This quantity is shown in the \fref{interferometer}(b), we can see that the condition of zero transmission probability is obtained if $\theta=\frac{\pi}{2}$ and $k_\text{SO}L=\frac{\pi}{2}$. 
%
%
\begin{figure}[!t]
\begin{center}
\includegraphics[width=\columnwidth]{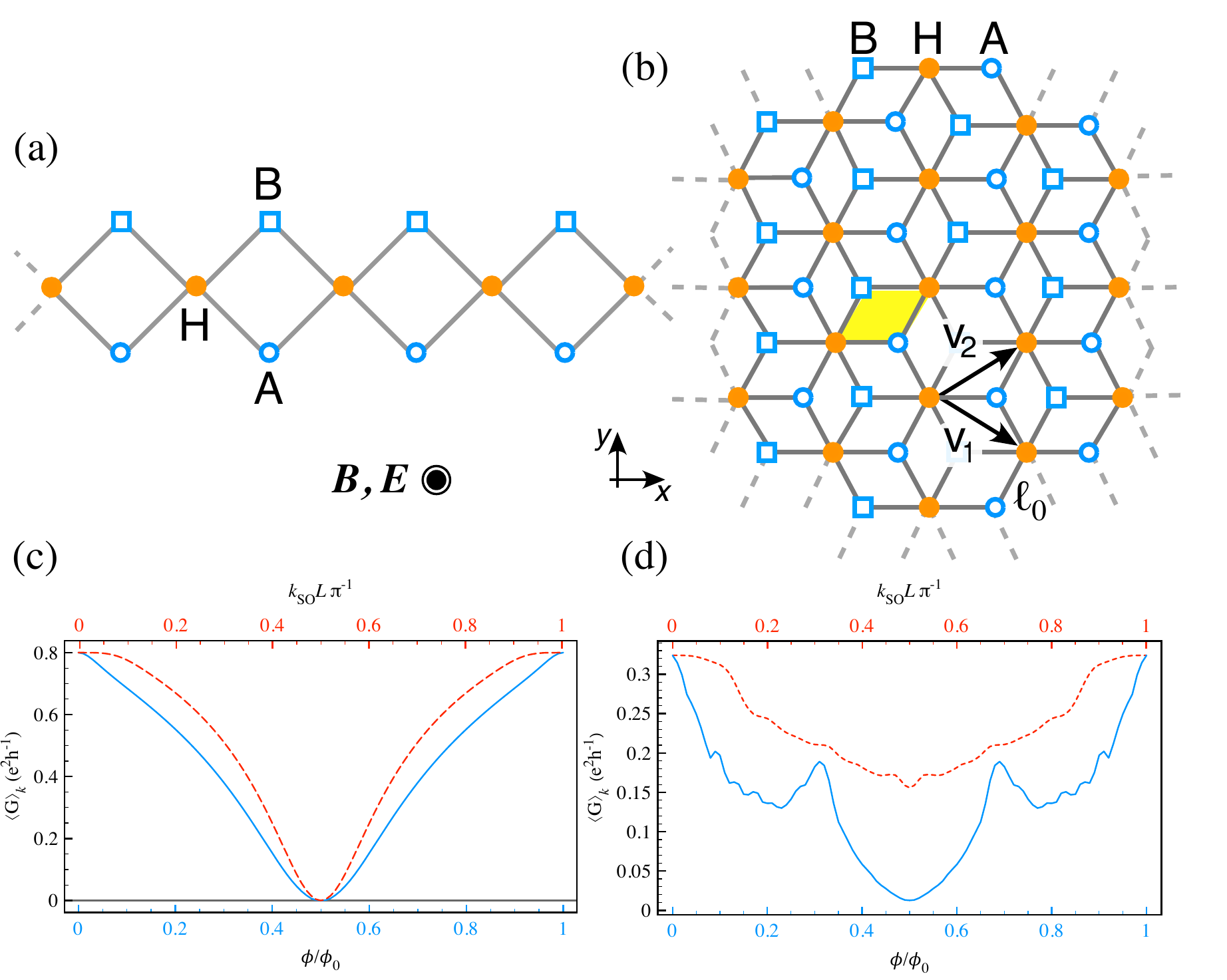}
\caption{\label{lattices}(Colours online) (a) Sketch of diamond chain, this is the one-dimensional version of the $\mathcal{T}_3$ lattice. (b): Sketch of the $\mathcal{T}_3$ lattice, it is a Bravais lattice with a unit cell containing three inequivalent sites: A, B and H, the latter has coordination number higher that the other two~\cite{Sutherland:1986}. (c): averaged conductance for the diamond chain as a function of the magnetic flux ratio (lower $x$-axis, blue solid-line) and as a function of the dimensionless RSOI (upper $x$-axis, red-dashed line) (Courtesy of Ref.~[\onlinecite{Bercioux:2005}]). (d): averaged conductance normalised to the number of leads for the $\mathcal{T}_3$ lattice as a function of the magnetic flux ratio (lower $x$-axis, blue solid-line) and as a function of the dimensionless RSOI (upper $x$-axis, red-dashed line) (Courtesy of Ref.~[\onlinecite{Bercioux:2005}]). In (c) and (d) \textbf{\emph{B}} and \textbf{\emph{E}} are the applied magnetic field and the electric field, respectively.}
\end{center}
\end{figure}
%
%
To summarise, while in the case of a U(1) gauge field we can obtain destructive interference for each geometrical realisation of the interferometer [see \fref{interferometer}(a)], in the SU(2)  case a specific geometric arrangement of the loop is required. 

These effects can have interesting  implications in the transport properties of nontrivial quantum networks.
In the following, we shall focus on a very special type of lattice structure: the $\mathcal{T}_3$ lattice [c.f. \fref{lattices}(b)], whose unit cell contains three inequivalent sites~\cite{Sutherland:1986}. Its spectrum is composed of two dispersive bands that are equivalent to the one of the honeycomb lattice and a flat band at zero energy that is due to the presence of lattice sites with uneven coordination number~\cite{Bercioux:2009,Bercioux:2011}. In the presence of a perpendicular magnetic field --- when the ratio $\frac{\phi}{\phi_0}=\frac{1}{2}$ --- the spectrum reduces to three flat bands~\cite{Vidal:1998}. This  localisation effect is a consequence of the AB effect  taking place within each of the plaquette contained in the $\mathcal{T}_3$ lattice [in yellow in \fref{lattices}(b)]. The same effect is observed in its quasi-1D analog, the diamond chain [c.f.~\fref{lattices}(a)] for the $\frac{\phi}{\phi_0}=\frac{1}{2}$~\cite{Vidal:2000}. 

Can the same physics be driven also by the RSOI? The question has been addressed  by one of the authors using the method of quantum networks~\cite{Naud:2001}. Full localisation due RSOI is possible only in the diamond chain~\cite{Bercioux:2004} [\fref{lattices}(c)] and is forbidden, for the arguments illustrated above, in the case of the $\mathcal{T}_3$ lattice~\cite{Bercioux:2005} --- \fref{lattices}(d). In \fref{lattices}(c) and \ref{lattices}(d) we show the conductance of the diamond lattice and the $\mathcal{T}_3$ lattice, respectively, as a function of the magnetic flux and RSOI. We observe that in the case of the diamond chain \ref{lattices}(c), both effects induce a complete localisation equivalent to zero conductance. However, in the case of the  $\mathcal{T}_3$ lattice \ref{lattices}(d), only the magnetic field can induce (almost) complete localisation. The residual conductance comes from the conducting state along the lattice boundary due to the  Hall effect. However it can be reduced to zero by injecting electrons through the centre of the lattice~\cite{Vidal:2000r}. 
Signatures of this AB caging effect have been observed also experimentally~\cite{Abilio:1999,Naud:2001}. The diamond chain lattice model with RSOI and AB field has been also proposed as a system for producing a spin filter~\cite{Hatano:2007,Aharony:2008,Aharony:2011,Aharony:2013,Matityahu:2013}.

\subsection{Interference in quantum rings}
Mesoscopic quantum rings allow to have direct access to the phase of the electron wavefunction,  when
their size is smaller than the coherence length.
Interference effects have been observed in metal quantum rings many years
ago \cite{washburn:1992}.
Since electrons are spinful particles, the spin part of the wavefunction is influenced by the magnetic field
via the Zeeman term in the Hamiltonian.  A more subtle
effect arises when there is a magnetic field non-orthogonal to the
plane of the orbiting particle (\emph{e.g.} the effective magnetic field due to the RSOI) because, as a consequence of the
orbital motion, its spin dynamics is instantaneously governed by a
time-dependent Hamiltonian \cite{anandans, Saarikoski:2014}.  This time dependence ends
up in an extra phase acquired by the particle wavefunction which is
named after Berry~[\onlinecite{berry,anandan}], who put in foreground its topological properties
when the orbits are closed.

During the last years, the effects of RSOI on the AB oscillations have
been observed in semiconductor based quantum rings by several groups
\cite{Yau:2002,Nitta:2003,Meijer:2004,morpurgo:1998}.  As said before, in the
presence of both orthogonal magnetic field and RSOI, the total
effective momentum dependent  magnetic field is tilted with respect to the vertical
direction. The resulting Berry phase influences the interference
pattern.

\begin{figure}[!hb]
\begin{center}
\includegraphics[width=\columnwidth]{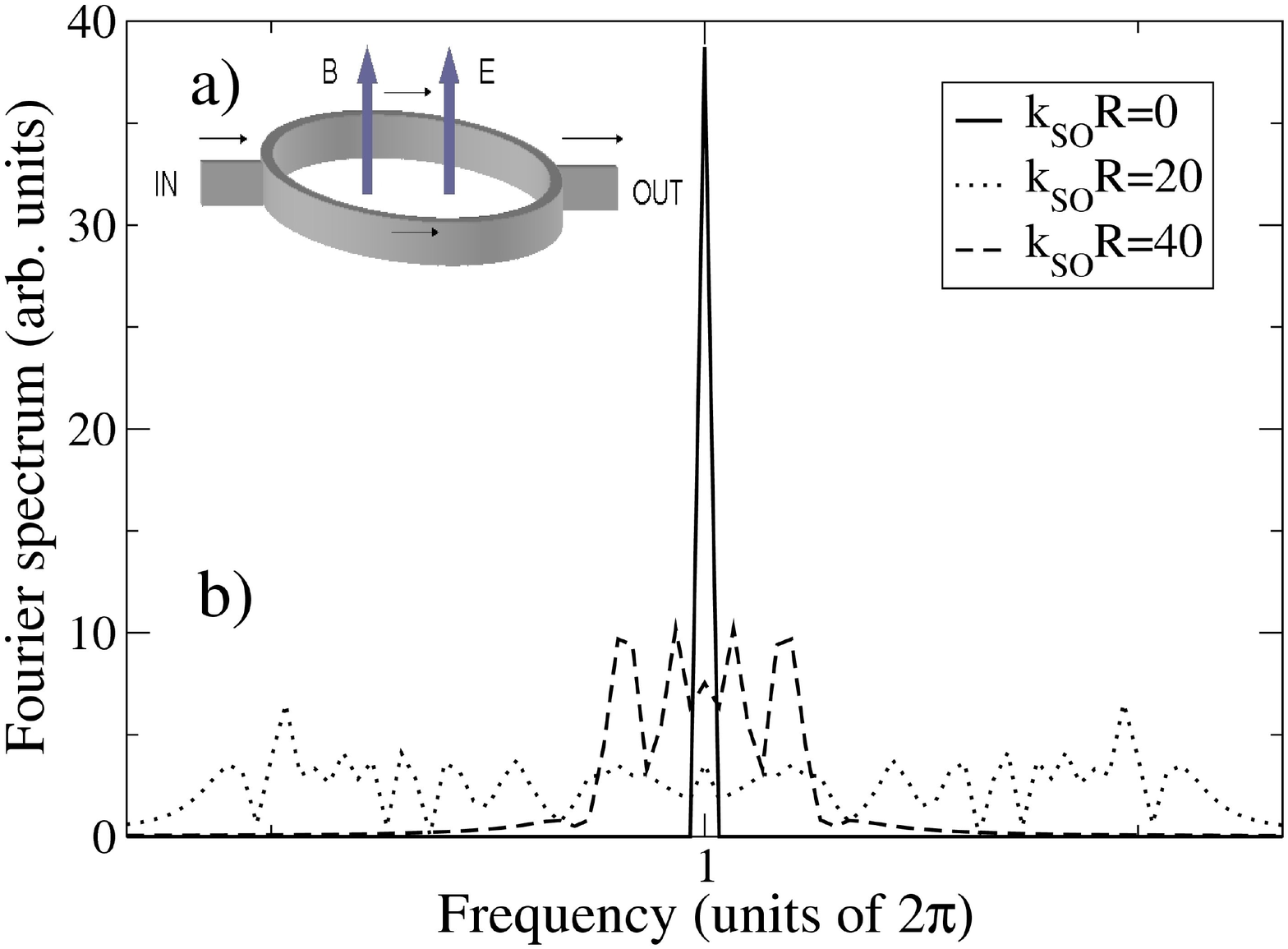}
\caption{\label{sidepeaks}{(Colours online) a): Sketch of a quantum ring. The magnetic field B drives the AB effect, while the electric field can tune the RSOI. b) Fourier transform of the magneto conductance.  By increasing ROI two satellite peaks rise close to the AB peak. They can be attributed to the Berry phase acquired by the electron during its motion.} Courtesy of Ref.~[\onlinecite{Capozza:2005}].}
\end{center}
\end{figure}

In Refs.~[\onlinecite{Capozza:2005,Lucignano:2007}] one of the authors studied the conductance and the spin transport 
in a quantum ring in the presence of RSOI and magnetic field,  accounting also for  dephasing at the
contacts. The AB resonance in the Fourier transform of the magneto-conductance 
displays satellite peaks due to the RSOI (see \fref{sidepeaks}) that have been experimentally observed~\cite{habib:2007}.
%
%
%
%
During the last years, several theoretical techniques have been employed to study quantum  rings. In Refs.~[\onlinecite{Loss:1990,Engel:2000}], an imaginary time path integral approach is developed to study the conductance of a strictly 1D quantum ring, and its conductance fluctuations in the diffusive limit. In
Ref.~[\onlinecite{tserkovnyak:2007}] a real time path integral approach is applied in the limit of negligible Zeeman splitting.  Several research articles 
discussed the conductance properties and the spin dependent transport of quantum rings  in the  1D ballistic limit, by means of a spin
dependent scattering matrix approach \cite{Molnar:2004,Frustaglia:2004,Bercioux:2005b,Citro:2006,Ramaglia:2006}. In the absence
of the magnetic flux, the conductance shows quasi-periodic oscillations in the RSOI strength, which can be modified by switching
the magnetic field on.  Numerical calculations \cite{Frustaglia:2004,souma:2004,wu:2006} have shown that in the 2D case there are only quantitative
modifications of the 1D results that do not qualitatively affect the physics. Thus, in the following, we will just focus on the 1D limit.

The model Hamiltonian, describing a half integer spin particle, in an orthogonal magnetic field and with RSOI~\cite{Meijer:2002},  reads
%
%
\begin{eqnarray}
\mathcal{H}[ \bm{p} , \bm{r} , \bm{S} ]  =
\frac{1}{2m}\left(\bm{p}+\frac{e}{c}\bm A_0\right)^2
-  \omega_\text{c}\: S_{z}+\mathcal{H}_\text{RSOI}
 \label{hamilt} \\
\mathcal{H}_\text{RSOI}= \frac{2\alpha}{\hbar ^2}\left[ {\bm z} {\times} 
\left({\bm p+\frac{e}{c} \bm A_0}\right) \right]{\cdot}
{\bm{S}}  \,, \nonumber
\end{eqnarray}
%
%
where $\bm{A}_0(\bm{r}) = \frac{B}{2} ( - y , x , 0 ) $ is the vector potential generating the uniform field $B$, normal to the ring
surface, and $\omega_\text{c} = \frac{g^*e B}{2mc}$ is the cyclotron frequency. We fix the vector potential in the symmetric gauge. We present, here,  fully general results as they depend on the ratio $\frac{\alpha}{\hbar \omega_\text{c} R}$ which can be tuned by acting on
$\alpha$.  We assume a single channel ring as a 1D circle of radius $R$, connected to two leads. Accordingly, the position of the particle on the ring is
parametrised by the angle~$\varphi$. The vector potential has just the azimuthal component $A_\varphi = \frac{\phi }{ 2\pi R}$, where $\phi$ is the
magnetic flux threading the ring.

In order to study the conduction properties of the ring, we need the propagation amplitude for an electron at energy $E_0$ entering the ring with spin
polarisation $s_0$ and leaving it with spin polarisation $s_f$. This is given by
%
%
\begin{equation}
A(s_f ; s_0 | E_0 )  =
\int_{0}^{\infty} \: \frac{d t_f}{\tau _0}  \:  \rme^{\rmi \frac{E_{0}
t_f}{\hbar}}  \:
\langle {\bm r}_{f} , s_{f} , t_{f}|{\bm r}_{0} , s_{0 } , t_{0} \rangle
 \,,
\label{ampclas1}
\end{equation}
%
%
where $\langle {\bm r}_{f} , s_{f} , t_{f}|{\bm r}_{0}, s_{0 } , t_{0} \rangle$ is the  amplitude for  a particle entering the ring  at the point $\bm{r}_0$ and  at the time $t_0$ with spin  polarisation $s_0$ to exit at the point $\bm{r}_f$ at the time $t_f$ with spin polarisation $s_f$.
Here, $\tau _0 =\frac{mR^2 }{2\hbar} $ is the time scale for the quantum motion.  In order to compute $\langle {\bm r}_{f} , s_{f} , t_{f}|{\bm r}_{0} , s_{0 } , t_{0} \rangle$, we adopt a path integral representation for the orbital part of the amplitude. Since we parametrise the orbital motion of the particle in terms of the angle $\varphi$, we provide the {appropriate} Lagrangian, ${\mathcal{L}}_\text{orb}$, as a function of $\varphi , \dot{\varphi}$.  It reads 
%
%
\begin{align}
\mathcal{L}_\text{orb}[\varphi(t), \dot \varphi(t),\bi \sigma]= &
\frac{m}{2}R^2\dot\varphi^2(t)-\frac{\phi}{\phi_{0}} \hbar
\dot\varphi(t) \nonumber \\
& + \frac{\alpha^2\:m}{2 \hbar^2}+ \frac{\hbar^2}{8 m
R^2} \,.
\label{lag}
\end{align} 
%
%
The last two contributions in equation \eqref{lag} are constants that come from the RSOI term, and the Arthurs  term, which is required when a path integration is performed in cylindrical coordinates. Since both contributions are constant, they can be lumped into the incoming energy $E_0\sim E_\text{F}$ and therefore they will be omitted henceforth.

By taking into account the spin degree of freedom, as well, we represent the propagation amplitude as 
%
%
\begin{align}
&\langle {\bi r}_{f}
, s_{f} , t_{f}|{\bi r}_{0} , s_{0 } , t_0 \rangle  = \langle
{\bi r}_{f} , s_{f} | \rme^{ - \rmi \int_{t_0}^{t_f } \: d t \: \mathcal{H} } | {\bi
r}_{0} , s_{0 } , t_0 \rangle \nonumber\\
&=\int_{\varphi ( t_0 ) =\varphi_0}^{\varphi ( t_f ) = \varphi_f} \!\!\!\!{\cal D} \varphi\: \rme
^{- \rmi \int_{0}^{t_f} d t \; \left[ \tau_0 \dot{\varphi} ^2 - q
\dot{\varphi}\right ]} \: \langle s _f | \bi{U}_\text{spin} ( t_f, t_0 ) |
s_0 \rangle \,,
\label{parte}
\end{align} 
%
%
where $q=\frac{\phi}{\phi_0}$. 
%
%
\begin{equation}
\bm{U}_\text{spin} ( t_f, t_0 ) = {\bf T} \exp \left [ -\frac{\rmi}{\hbar} \:
\int _{t_0}^{t_f} d\tau \: \bm{H} _\text{spin} (\tau) \right ] \,.
\end{equation}
%
%
is the full  spin  propagator, {$\bf{T}$ is the time ordering operator}, and $ \bm{H} _\text{spin} (t)$ is the spin Hamiltonian that reads 
%
%
\begin{equation}
\bm{H}_\text{spin} ( t ) = \frac{\hbar}{2}
 \left( \begin{array}{cc}  \omega_\text{c} & 
 2 \gamma  \dot\varphi  \rme^{ - \rmi \varphi ( t ) } \\
2 \gamma  \dot\varphi \rme^{  \rmi \varphi (t)  } &
-  \omega_\text{c}  \end{array} \right)
\,, 
\label{three}
\end{equation}
%
%
with $\gamma = \frac{2\alpha\tau _0}{\hbar R}$.

\begin{figure}[!htp]
    \centering \includegraphics[width=0.6\columnwidth]{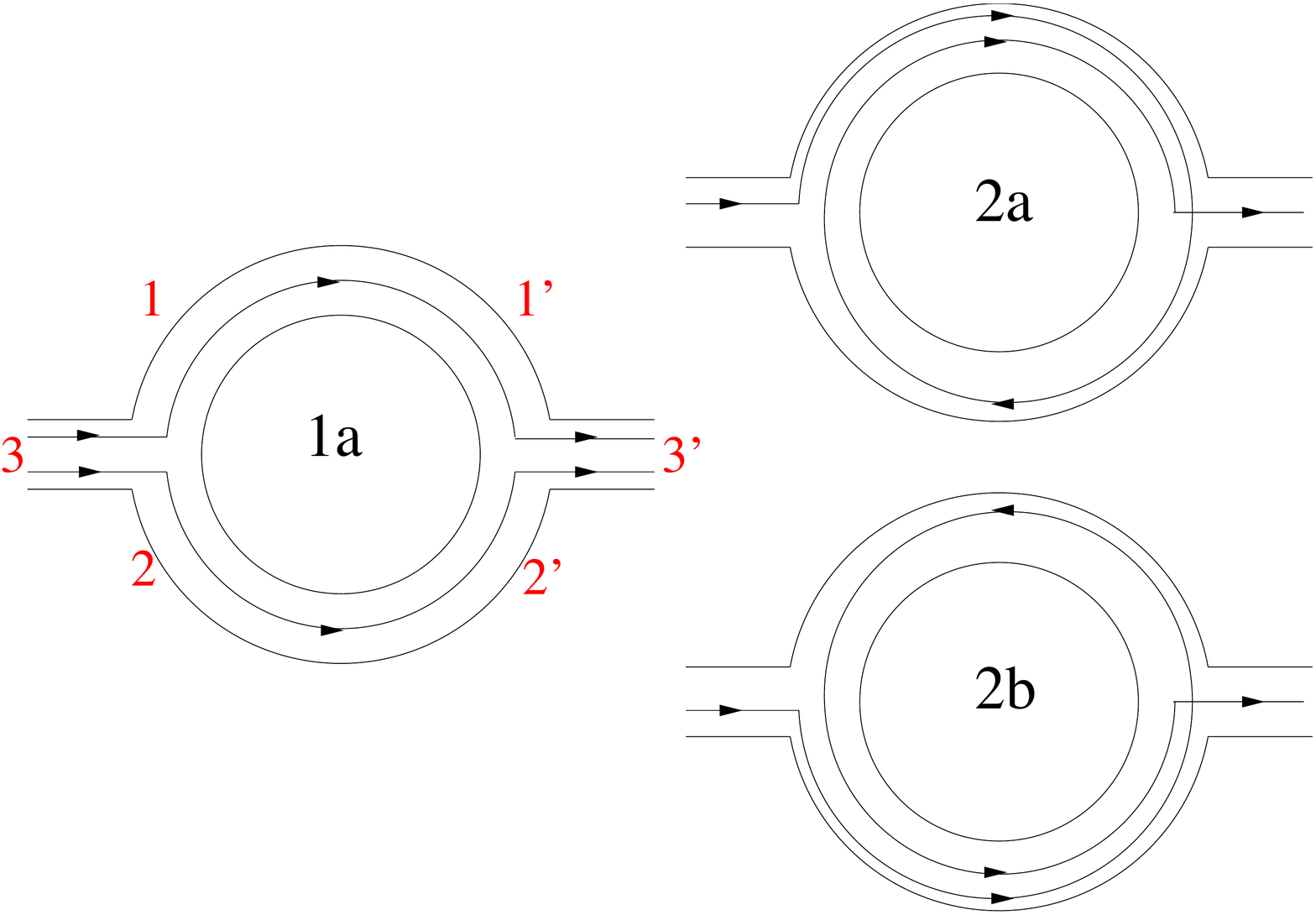}
 \caption{(Colours online) First and second order paths included in the calculation of the transmission amplitude across the ring, from left to right, including forward scattering only. Numbers $1,2,3(1',2',3')$ in \fref{noWLpaths}a refer to the labelling of the terminals in Eq.(\ref{matr}). Courtesy of Ref.~[\onlinecite{Lucignano:2007}].}
\label{noWLpaths}
\end{figure}

For a ring device, at each lead one has to take into account three possible scattering processes, consistently with the conservation of the
total current. This is  described in terms of a unitary $\mathcal{S}-$matrix that, when the two arms are symmetric, is given by
%
%
\begin{equation}
\mathcal S=\left(
\begin{array} {c c c}
         -\frac{1}{2} (1+ r) & \frac{1}{2} (1- r)          & \sqrt{\frac{1}{2}(1-  r^2)}\\
          \frac{1}{2} (1- r) &-\frac{1}{2} (1+  r)          & \sqrt{\frac{1}{2}(1-  r^2)}\\
 \sqrt{\frac{1}{2}(1-r^2)}  & \sqrt{\frac{1}{2}(1-  r^2)}  &    r
\end{array}\right)
\label{matr}
\,.
\end{equation}
%
%
The numerical labelling of the $\mathcal{S}$-matrix elements referring to the three terminals of each contact fork, are explained in \fref{noWLpaths}(1a).  Assuming, for simplicity, that the scattering matrix is the same for both leads, equation \eqref{matr} will hold both at the left-hand lead,
and at the right-hand lead of the ring.
%
%
%
%

\begin{figure}[!htp]
    \centering
    \includegraphics[width=0.6\columnwidth]{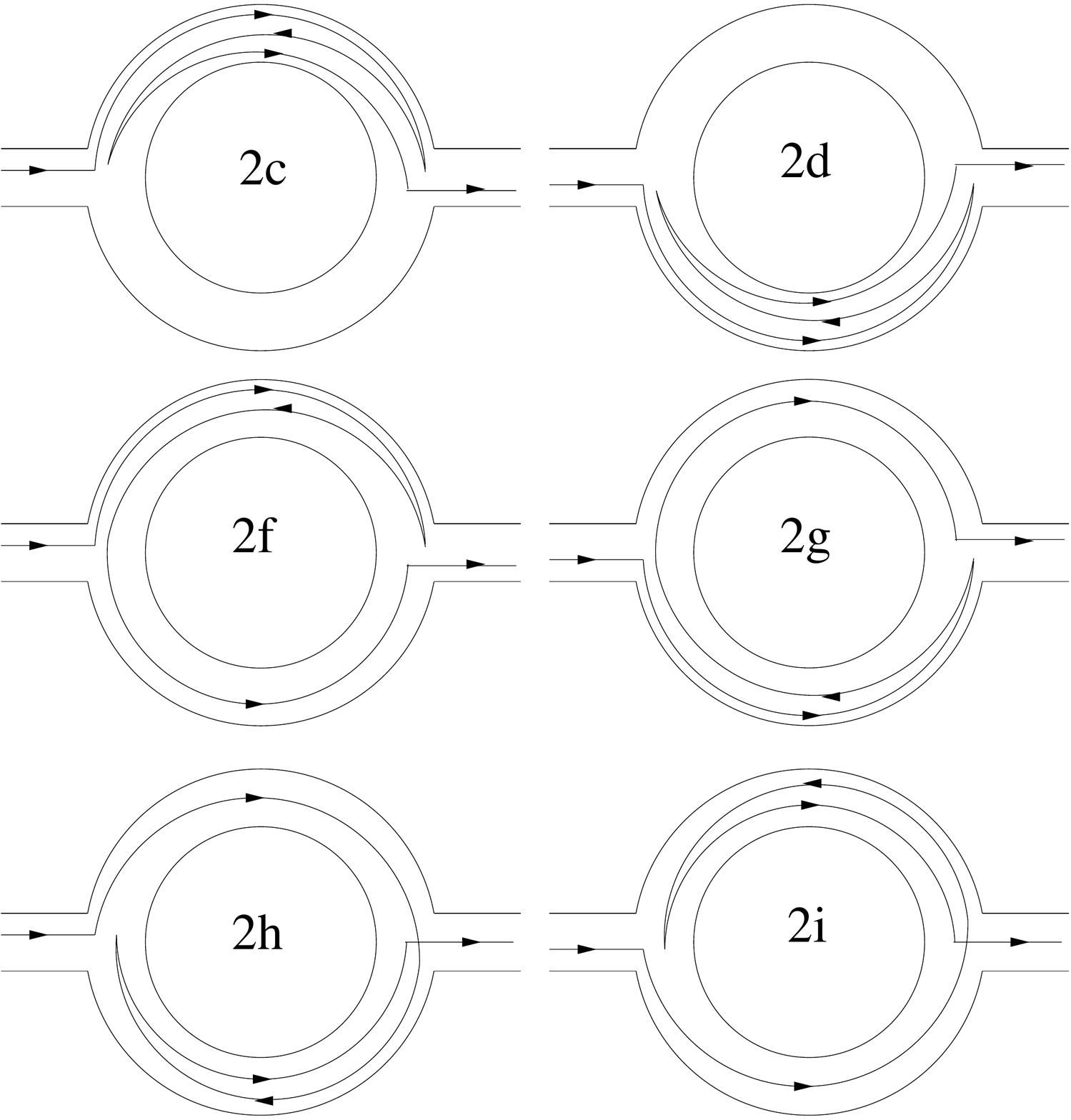}
    \caption{Second order paths of the transmission amplitude from left to  right including backscattering at the leads. Paths $(2f)$ and
  $(2h)$, as well as $(2g)$ and $(2i)$ contribute to the weak  localization corrections.  Courtesy of Ref.~[\onlinecite{Lucignano:2007}].}
\label{WLpaths}
\end{figure}

In particular, $\mathcal S_{3,3}=r$ is the reflection amplitude for a wave coming from the left lead, $\mathcal S_{1(2),1(2)}=-\frac{1}{2}(1+r)$ the reflection amplitude for a wave incoming from the upper/lower arm, $\mathcal S_{1(2),2(1)}= \frac{1}{2} (1-r)$ is the transmission amplitude from the upper (lower) to the lower (upper) arm and $\mathcal S_{1(2),3}=\mathcal S_{3,1(2)}=\sqrt{\frac{1}{2}(1-r^2)}$ is the transmission amplitude from the upper/lower arm to outside of
the ring.  As the ring is assumed to be symmetric, the same scattering matrix applies to the right lead where we indicate with primed numbers the three terminals as shown in \fref{noWLpaths}.
In \fref{noWLpaths}  we show the simplest paths of the electrons in the ring including only forward scattering at the contacts. More involuted paths arise if we account also for backscattering processes in which the electron can get backscattered within the same ring's arm from which it is coming. For instance, the paths $(2f)$ and $(2h)$, as well as $(2g)$ and $(2i)$ in \fref{WLpaths}, include looping in opposite directions around the ring. Interference between clockwise and counterclockwise windings leads to WL corrections. We denote these corresponding paths  --- including also $(2c)$ and $(2d)$ --- as ``reversed paths''. In our
approach, all order paths are numerically generated up to the convergency and the $\mathcal{S}-$matrix \eqref{matr} is implemented in the numerical algorithm.
%
%
%
%
Within Landauer's approach, the conductance ${\cal{G}}$ is given by
%
%
\begin{equation}
\mathcal{G}=
\frac{e^2}{\hbar}
\sum_{s,s'}
 \left | \mathcal {A} (s;s'|E_\text{F})\right|^2
\,. 
\label{ec.1}
\end{equation}
%
%
Here we will  consider the dependence on the external magnetic field $\frac{\phi}{\phi _0}$ and on the RSOI  strength $k_\text{SO} R$ both in the absence and in the presence of dephasing at the contacts. To make the model more realistic, we allow for higher order looping of the electron within the ring.  In Ref.~[\onlinecite{Capozza:2005}], only the paths of the kind of figures \ref{noWLpaths}(2a) and \ref{noWLpaths}(2b) were included.  Following Ref.~[\onlinecite{Lucignano:2007}] we consider here also the paths of the kind of \fref{WLpaths} in which the electron can be backscattered into the ring.  We  use here $r=0$ in the scattering matrix between the arms and the leads, which means that no back-reflection in the incoming lead is present. 

\begin{figure}[!htp]
    \centering \includegraphics[width=0.9\columnwidth]{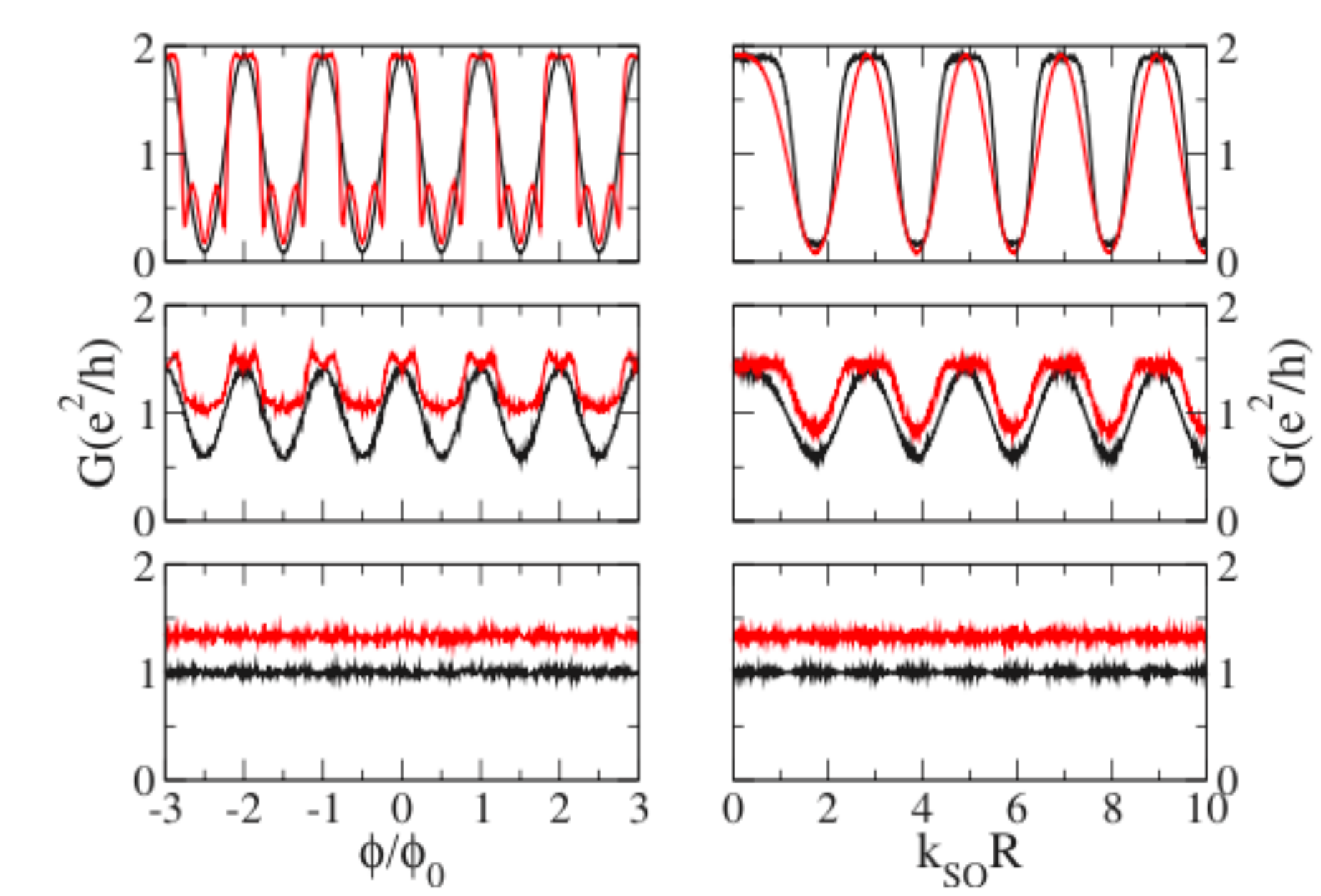}
\caption{(Colours online)Conductance as a function of $\phi/\phi_0$ {\sl (left panels)} and $k_\mathrm{SO}R$ {\sl (right panels)} for ideal {\sl
(black curves)} and realistic {\sl (red curves)} contacts. An increasing amount of dephasing at the contacts is also included: {\sl from top to bottom:} $\zeta=\pi/3,\pi,2\pi$. Courtesy of Ref.~[\onlinecite{Lucignano:2007}].}  
\label{GvsBSO}
\end{figure}

The dephasing due to diffusiveness in the contacts is accounted by adding a random phase $z\in (-\zeta,\zeta ) $ for each scattering at the leads.  In \fref{GvsBSO}, we report the conductance $\mathcal{G}$ as a function of $\frac{\phi}{\phi_0}$ , with $k_{\text {SO}}R = 0$ (left column) or as a function of  $k_\text{SO} R$ with $\frac{\phi}{\phi_0}=0$ (right column). These are averaged over $N=1000$ realisations of dephasing, and plotted for increasing window of phase randomness ($\zeta = \pi/3,\pi, 2\pi$ from top to bottom). The black curves refer to ideal contacts (including only the paths of \fref{noWLpaths}) while the red curves refer to realistic contacts (including also the paths of \fref{WLpaths}).

The ring is rather insensitive to small dephasing at the contacts however, by increasing the amount of dephasing (middle and bottom left panels in \fref{GvsBSO})  we find that the sensitiveness is larger  in the case of realistic contacts. This is due  to the fact that for realistic coupling, the electrons in the ring can experience higher order paths, since it scatters with the leads many times.  

%
%
%
%
%
In the right panel of \fref{GvsBSO}, we plot the \emph{dc}-conductance as a function of  $k_\text{SO}R$ at $\phi =0 $ for both ideal contacts and realistic contacts  (black and red lines in each box), with an increasing phase randomisation  (boxes from top to bottom with $\zeta = \frac{\pi}{3} ,\pi, 2\pi$), averaged over $N=1000$ disorder realisations. In the case of ideal contacts and little dephasing (top right panel black curve), we observe again quasi periodic oscillation of the conductance reproducing, the localisation conditions at the expected values of $k_\text{SO}R$ \cite{Frustaglia:2004,Molnar:2004,Bercioux:2005b,Capozza:2005,Lucignano:2007}.  When including higher order processes, interference effects give rise to a slightly different pattern.  In the case of realistic contacts, we note that the conductance of the ring is seriously  affected by dephasing. Indeed, large dephasing gives rise to random oscillations that  are not averaged out ---  thus washing  out the conductance oscillations. The effect takes place  for $\zeta\sim\pi$  when time reversed paths are included, in contrast to $\zeta \sim 2\pi$ when  the time reversed paths are  absent.  As regular magneto-conductance oscillations are experimentally observed \cite{Nitta:2003,Meijer:2004,Yau:2002,morpurgo:1998} with  little percentage of contrast between maxima and minima, we conclude that, in real samples, dephasing is ubiquitous.

\section{Related problems}\label{related}
\subsection{Spin-Hall effect}\label{SHE}
The SHE is a phenomenon, {associated} to SOI --- it can be used to electrically generate or detect spin currents in non-magnetic systems. This effect has been observed both in metallic and semiconductor  {systems} (for a review see Ref.~[\onlinecite{Jungwirth:2012,vignale10}]). In this review, we will mainly focus on semiconductors, in which RSOI plays a major role. 
The SHE predicts that  an unpolarised electric current can generate a transverse spin current, whose spin is perpendicular to the plane of the two currents. 
It was predicted in two pioneering articles  by Dyakonov and Perel in 1971~[\onlinecite{Dyakonov:1971a,Dyakonov:1971b}]. It is  a consequence of the Mott scattering of electrons on unpolarised impurities, which results in spatial separation of electrons with opposite spins. It is closely related to the anomalous Hall effect  --- Hall effect in ferromagnetic materials --- originally observed by Hall himself~[\onlinecite{Hall:1881}]  and later explained by  Karplus and Luttinger~[\onlinecite{Karplus:1954}] and Nozieres and Lewiner~[\onlinecite{Nozieres:1973}]. It does not require magnetic field nor magnetism, in other words it does not require broken time reversal symmetry (for a review see [\onlinecite{nagaosa:2010}]). 
The SHE was hardly investigated till Hirsch~[\onlinecite{Hirsch:1999}] and Zhang~[\onlinecite{Zhang:2000}] proposed  such phenomenon  to the attention of the spintronics community. 
Three main mechanisms have been proposed to contribute to the SHE: namely the spin dependent band structure of the material (the so called intrinsic mechanism) and the impurity scattering mechanisms, \emph{i.e.} the ``skew scattering" and the ``side jump", conventionally addressed as extrinsic mechanisms.
The extrinsic mechanisms are the same responsible of the anomalous Hall effect \cite{nagaosa:2010}: the skew scattering originates from the different scattering angle due to a spin-orbit impurity, depending on the electron spin and angular momentum;  while the side jump reflects the shift in the trajectory of the backscattered electrons originating from the anomalous velocity  operator in SOI systems \cite{crepieux:2001}.
The intrinsic mechanism was proposed by Murakami \emph{et al.}~[\onlinecite{Murakami:2003}] and by Sinova \emph{et al.}~[\onlinecite{Sinova:2004}], it resorts on the DSOI and RSOI in 2D III-V GaAs quantum wells. This mechanism does not explicitly depend on  
impurities, however these cannot be neglected as they are absolutely essential to the establishment of the steady state current. In principle, the spin Hall current can be calculated using the Kubo formula. However, inclusion of vertex corrections due to impurities, in the linear SOI model leads to a vanishing spin Hall conductivity \cite{Raimondi:2005,Dimitrova:2005} in striking contrast to earlier calculations not including vertex corrections \cite{Sinova:2004}. More complicated models (explicitly carried out for transition metals) \cite{Tanaka:2008} overtake this problem, giving rise to a finite spin Hall conductivity in reasonable agreement with the experiments. 
Some proposals require the presence of a magnetic field to give rise to a finite spin Hall current even in the presence of linear SOI and disorder \cite{Engel:2007,Lin:2006,Milletari:2008,Lucignano:2008,Raimondi:2006,Gorini:2008}.
Among the first successful experiments we here highlight the work by Wunderlich \emph{et al.}~[\onlinecite{Wunderlich:2005}] that used coplanar \emph{p-n} diodes to detect circularly polarised electroluminescence at opposite edges of the spin Hall bar, and the work by Kato \emph{et al.}~[\onlinecite{Kato:2004}] that employed  a magneto-optical Kerr microscope to scan the spin polarisation across the Hall bar. The former ascribed their finding to the intrinsic mechanism whereas the latter to the extrinsic one.
Since that time several interesting experiments performing optical measurements for the spin detection in  the intrinsic and extrinsic SHE have been performed \cite{matsuzaka09,nomura05,sih05,sih06,stern06,stern07,stern08,zhao06,okamoto14}. 
A detailed description of this physics is out of the purpose of this reviews and we refer the readers to more specific review articles on the SHE for more information~\cite{Jungwirth:2012,vignale10}.

\subsection{Diffusive limit: weak anti-localization}\label{WAL}
In the previous sections we have mainly described the physics of ballistic mesoscopic systems, \emph{i.e.} solid state devices in which the lateral size of the sample $L$ is smaller than the coherence length $L_\phi$ and of the elastic mean free path $\ell_e$. In this regime, impurity scattering can be safely disregarded, and the ballistic motion of electrons well describes the physical scenario. However, in the mesoscopic limit, it may happen that the lateral size of the sample $L$, while being still smaller (or of the order of) than the  coherent length of the material $L<L_\phi$, it is larger than the elastic mean free path: $L>\ell_e$.  Impurity scattering cannot be neglected and the electrons scatter in the sample undergoing a quantum diffusive dynamics.
In a 2DEG at low temperature, quantum corrections to the conductance  are mostly due to interference effects between two electrons scattered by the same impurity. These propagate along the same closed trajectory but in opposite directions  --- usually known as Cooperon correction. This process has its more spectacular manifestation in the so called  WL (see Ref.~[\onlinecite{Akkermans:2007,Altshuler:1980,Kramer:1993}]): it increases the effective scattering cross-section and therefore leads to a suppression of the conductivity.  In the presence of a magnetic field $\bm{B}$ perpendicular to the electron plane, the two waves acquire a phase difference $\Delta\varphi=2\frac{\phi_{B}}{\phi_0}$, where $\phi_{B}$ is the magnetic flux through the area enclosed by the electron trajectory, therefore quantum interference effects  are suppressed and the conductance increases to  the classical limit. Such a modification  in conductivity, as a function of the magnetic field, is known as positive magneto conductivity (sometimes called negative magnetoresistance).
This quantum correction to the resistivity  can be  explicitly calculated evaluating the Cooperon diagram. For a 2DEG in perpendicular magnetic field 
%
%
\begin{equation}
\Delta \sigma(B) \propto \int_0^\infty\!\!\!\!\!\! \mathrm{d}t\: Z_c(t,B) \!\!\left[\rme^{-\frac{t}{\tau_\phi}}-\rme^{-\frac{t}{\tau_e}} \right]\,,
\end{equation}   
%
%
where $Z_c(t) = \frac{\phi_B}{\phi_0}\sinh^{-1}\left(\frac{4 \pi Dt B}{\phi_0}\right)$, where $\tau_\phi$ and $\tau_e$ are phase coherence time and electron relaxation time{and $D$ is the 2D diffusion coefficient}, respectively. 
In the limit of very weak magnetic field ($B\ll\frac{\phi_0}{8\pi L_\phi^2}$) the magnetoresistance results to be:
%
%
\begin{equation}
\Delta R(B) \propto - \frac{\Delta \sigma (B) }{\sigma_0^2}\,,
\end{equation}
%
%
thus negative, and the resistivity correction can be shown to be 
%
%
\begin{equation}
\Delta \sigma (B) -\Delta \sigma (0) \propto \left(\frac{B}{B_\phi}\right)^2\,,
\end{equation}
%
%
where $B_{\phi} = \frac{\phi_0}{8 \pi L_\phi^2}$ is a characteristic magnetic field corresponding to a flux quantum through an area of the order of $\pi L_{\phi}^2$. Measuring the negative magneto-resistance by applying a  weak magnetic field is a very elegant way to probe weak localization correction to the conductance.

In the presence of SOI the   WL  correction changes its sign and the magnetoresistance results to be positive. Indeed, Cooperon correction is multiplied by a factor 
%
%
\begin{equation*}
\langle Q_\mathrm{SOI}(t) \rangle = \frac{1}{2} \left[3 \rme^{-\frac{t}{\tau_\mathrm{SOI}}} -1\right]\,,
\end{equation*}
%
%
where $\tau_\mathrm{SOI}$ sets the time scale of the SOI.  For strong SOI, the exponential factor can be neglected at any time  $t$, thus the factor $\langle Q_\mathrm{SOI} \rangle$ changes its sign and tends to $-\frac{1}{2}$. It results in a change of the sign of conventional weak localization correction. This phenomenon,  theorised in Ref.~\cite{Hikami:1980},  is called WAL. It was first observed in metal films in the presence of spin-orbit impurities \cite{Bergman:1984,Bergman:1982} and has been intensively studied in the recent past, both in 2DEGs  \cite{Knap:1996,Koga:2002,miller:2003} and in large quantum dots \cite{Zumbuhl:2002,Aleiner:2001} made out of III-V semiconductors in the presence of RSOI [c.f. for instance \fref{alpha:variations}(b)]. Recently, weak WAL measurements have been used to quantify the RSOI strength in oxide interfaces~\cite{caviglia:2010} and TIs~\cite{Bardarson:2013}.

\subsection{Quantum wires with RSOI, superconductivity and magnetic field: {The quest for Majorana Fermions}\label{MFsec}} 
In his seminal article on the symmetric theory of electron and positron \cite{Majorana:1937}, Ettore Majorana predicted the existence of selfadjoint fermions, \emph{i.e.} half integer spin particles being their own antiparticle, as real solutions of Dirac equation. Since the first Majorana's prediction, the quest for Majorana Fermions (MFs) as elementary particles, in the high energy physics, is still open. Experiments on the neutrinoless double beta decay should unveil whether neutrinos are MFs or not, but to the date there is no answer to this question. 

On the contrary, electrons in solid, under particular circumstances could recombine with holes --- their relative anti-particles in the solid state language --- in order to form self-adjoint excitations, \emph{i.e.} Majorana quasiparticles.

Thus MFs \cite{Kitaev:2001,Kitaev:2003} can appear as quasiparticle excitations in solid state systems. For instance, they are expected to show up as boundary states of the Kitaev model:  a toy-model Hamiltonian describing 1D spineless $p$-wave superconductors. 

Despite the fact that ``solid state'' MFs  are not elementary particles, there is strong excitement in the condensed matter community, and many efforts are being devoted to their theoretical and experimental understanding.  
Of course, finding a novel  elementary excitation in solids would have a fundamental  significance on its own, however, in the case of MF there is strong interest also for its possible applications. Indeed MFs, due to their non-Abelian statistics \cite{Read-Green:2000,Ivanov:2001,Stern:2004,Alicea:2011} can be considered as building blocks for topological quantum computation and for other alternative  decoherence free quantum computational schemes \cite{Nayak:2008}.

Several mechanisms and devices have been proposed to isolate and detect MFs in meso/nano scaled devices (c.f. Refs.~[\onlinecite{Alicea:2012,Beenakker:2013}] for exhaustive reviews on the topic). However, to the date, while many efforts are focused on hybrid TI superconductor heterostructures \cite{veldhorst:2012, galletti:2014_1,galletti:2014_2}, (one of) the most promising device resorts on the use of III-V semiconducting nanowires with strong RSOI, in the proximity with conventional superconductors and in the presence of a magnetic field along the wire.
This mechanism, early proposed in Refs.~\cite{oreg:2010,lutchyn:2010} has been later experimentally explored in Refs.~[\onlinecite{Heiblum:2012,Kouwenhoven:2012,Churchill_PRB:2013}]. While strong hints of MF physics have been highlighted in Ref.~[\onlinecite{Kouwenhoven:2012}], other recent articles~[\onlinecite{Linder:2010,Churchill_PRB:2013,DeFranceschi:2012,Liu:2012,Kell:2012,SunKuei:2015}] offer alternative interpretation of the observed zero bias anomaly  in terms of Kondo physics.  Therefore an unambiguous ``smoking-gun''  experimental proof of the presence of MFs still has to come.

{In the following,} we will describe the theoretical background supporting the physics of MFs in quasi-1D nano wires with RSOI, magnetic field and proximity induced superconductivity. After that, we  briefly review a recent experimental result.

For sake of simplicity we assume a single active channel in the nanowire and set $\hbar=1$. The nanowire is along the $x$ direction, therefore the transverse dynamics, along $y$ direction, can be neglected. The Hamiltonian \eqref{hamiltonian_wire} describing the wire can be simplified as follows:
\begin{widetext}
\begin{eqnarray} \label{bdg1}
\mathcal{H}_\mathrm{MF} & = \sum_{\sigma \sigma'} \int_{-\infty}^{\infty}  \psi^\dagger_\sigma(x) \left(\frac{p_x^2}{2m^*}-\alpha s_y p_x + g \mu_B B s_x -\mu \right)  \psi_{\sigma'}(x) +\left (\Delta   \psi^\dagger_\uparrow(x)  \psi^\dagger_\downarrow(x) + \mathrm{h.c.} \right)\:dx  
\end{eqnarray}
%
%
\end{widetext}
where we have also added a Zeeman term due to the magnetic field $B$ along the wire axis and the superconducting pairing potential $\Delta$.
The operator $\psi^\dagger_\sigma(x)$ creates an electron of spin $\sigma$ at the position $x$.
One can notice that the Hamiltonian \eqref{bdg1} is not that of a 1D spineless $p$-wave superconductor, indeed we can easily recognise a local $s$-wave superconducting pairing and spinful electrons. {However, by analysing the action of the various interaction terms in \eqref{bdg1}, we can show how to obtain an optimal choice of the parameter regime that would map \eqref{bdg1} into the effective Hamiltonian of a 1D spinless $p$-wave superconductor.}

\begin{figure}[!h]
	\centering
	\includegraphics[width=0.9\columnwidth]{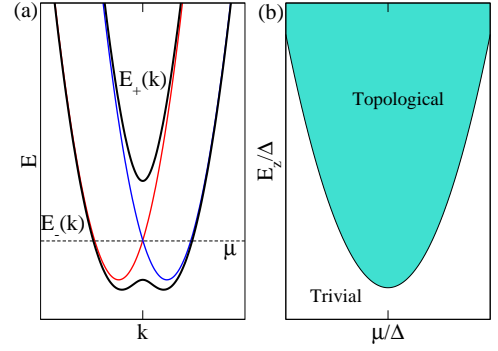}
		\caption{(Colours online) (a)  Free electrons band splitter by the simultaneous application of RSOI and Zeeman splitting.  
		(b) Phase diagram of the model Hamiltonian~\eqref{bdg2}.
		 \label{spectrum_Majo}}
\end{figure}

We choose $\Psi(x) = \left[u_\uparrow (x),u_\downarrow (x),v_\downarrow (x),-v_\uparrow (x)\right]$ as a Nambu spinor --- where $u/v$ are quasi electron$/$hole wave functions of momentum $k$ \cite{deGennesbook}. Thus the Bogolubov--de Gennes (BdG) Hamiltonian     \eqref{bdg1} in the $\Psi(x)$ basis, can be recast in the simple matrix form:
%
%
 \begin{equation} \label{bdg2}
\mathcal{H}_\mathrm{BdG} (k) =  \left(\frac{k^2}{2m^*}+\mathrm{i} \alpha k s_y  -\mu\right)\tau_z + E_z s_x   +\Delta\tau_x 
\end{equation}
%
%
where the $\tau_i$ Pauli matrices act in the Nambu space, while $s_i$ act in the spin space and we have introduced the Zeeman energy $E_z $.
In the absence of magnetic field and superconductivity, the conventional spin-degenerate parabolic band dispersion splits because of RSOI, into two parabolas crossing at $k=0$  {[c.f. \fref{spectrum2d}(e)]}. The Zeeman coupling, splits the two bands at $k=0$ giving rise to the two bands $E_+(k)$ and $E_-(k)$ in \fref{spectrum_Majo}(a).
If the Fermi energy lies between the two bands, and the temperature is low enough that the higher band cannot be thermally populated, the system is effectively a 1D spinless system. The spin degree-of-freedom has been effectively quenched by the simultaneous action of the RSOI and Zeeman splitting. Only a single  pseudo spin degree-of-freedom  ``$-$" is involved here.
In addition we have the $s$-wave superconducting pairing. Projecting the full Hamiltonian onto the $E_-(k)$ band  we can write down an effective Hamiltonian 
%
%
\begin{equation}
\mathcal{H}_\mathrm{P} =  \sum_k \left(E_{-}(k) c^\dagger_k c_k + \Delta_-(k)c^\dagger_k c^\dagger_k + \mathrm{h.c.} \right)
\end{equation}
%
%
where $\Delta_-(k) = \mathrm{i} \alpha k_x \Delta(\alpha^2 k_x^2 + E_z^2)^{-1/2}$ is an effective superconducting pairing with the desired $p$-wave symmetry. 
Thus, the present problem is isomorphic to the Majorana wire considered by Kitaev~[\onlinecite{Kitaev:2001}]. That guarantees the presence of MFs as boundary excitations.
However in the spirit of a review article, here we will not provide an explicit proof of the presence of MFs as boundary excitations of the wire, as it can be inferred by the isomorphism to Kitaev model and can also be  found in Refs.~[\onlinecite{oreg:2010,lutchyn:2010}] but we will provide  the reader with simple symmetry arguments to qualitatively justify this result.
%
%
%
%

Let us start by commenting  on the spectrum. By squaring twice Hamiltonian~\eqref{bdg2} we obtain the dispersion relation for the two bands
%
%
\begin{align*}
E^2_{\pm}(k,\Delta) =& E_z^2+\Delta^2+\xi_k^2+(\alpha k)^2 \\
&\pm 2\sqrt{(E_z\Delta)^2+(E_z\xi_k)^2+(\alpha k \xi_k)^2}
\end{align*}
%
%
where $\xi_k = k^2/2m-\mu $. The pairing $\Delta$  plays two crucial roles. It opens a gap at the outer wings of the dispersion, where the Zeeman field is unimportant, and modifies the gap forming near $k = 0$. The former role eliminates the possibility of high-momentum gapless excitations, thus leaving only the chiral states near $k = 0$ as low energy excitation.  The latter role allows us to tune the topological phase transition essential for isolating MFs.
One can easily verify that the Hamiltonian \eqref{bdg2} satisfies  particle-hole symmetry:
%
%
\begin{equation}
\Xi \mathcal{H}_\mathrm{BdG} (k) \Xi^{-1} = \mathcal{H}_\mathrm{BdG} (-k)\,.
\end{equation}
%
%
Hence, for each eigenstate of positive energy $\mathcal{H}_\mathrm{BdG}\psi = E \psi$ it exists a corresponding eigenstate $\Xi \psi$ of opposite energy $\mathcal{H}_\mathrm{BdG} (\Xi\psi) = -E (\Xi \psi)$. A linear combination $\gamma = \psi+(\Xi\psi)$ is of course a selfadjoint Fermionic operator, \emph{i.e.} a MF, however it is not an Hamiltonian eigenvector, unlike $E=0$~\cite{Chamon:2010}.

 %
 %
\begin{figure}[!t]
	\centering
	\includegraphics[width=\columnwidth]{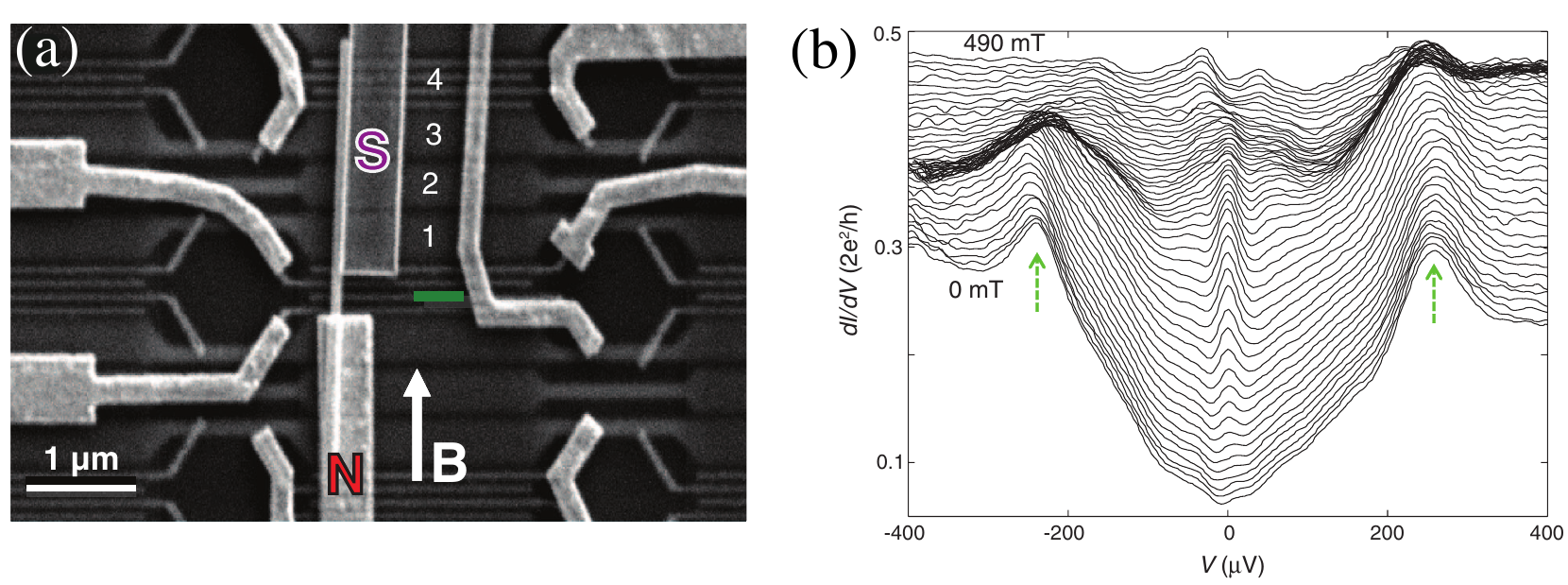}

		\caption{(Colours online)  (a): Experimental device. Electrons from a normal metal are injected in the InSb nanowire which is in contact with NbTiN  superconductor. (b): Differential conductance traces as a function of the magnetic field. As soon as the topological criterion is satisfied, a zero bias peak appears.  \label{Kouwenhoven_dIdV}Courtesy of Ref.~[\onlinecite{Kouwenhoven:2012}].}
\end{figure}
%
%
It can be shown that the presence of such zero energy selfadjoint eigenstate is indeed related to the sign of the {gap at $k=0$} 
%
%
\[
E_0=|E_+(k,\Delta)-E_-(k,\Delta)|_{k=0}=|E_z - \sqrt{\Delta^2+\mu^2}|
\]
%
%
between the two excitation branches $E_{\pm}(k,\Delta)$.  When the Zeeman energy $E_z$  closes such gap, we have a quantum phase transition between two different physical regimes. At $E_z >\sqrt{\Delta^2+\mu^2}$  the gap is a Zeeman gap while in the opposite case the gap is due to superconductivity.  Explicit construction of the zero energy excitation in the two cases shows that when the gap is magnetic, selfadjoint zero energy excitations, \emph{i.e.} MFs can appear at the two boundaries of the wire. The phase diagram of such system is shown in \fref{spectrum_Majo}(b).  
Out of the topological regime no MFs are expected, {but} the interesting physics of anomalous Josephson effect sets in \cite{Yokoyama:2014, Campagnano:2014}.

In Ref.~[\onlinecite{Kouwenhoven:2012}]   electrons are injected from a normal metal in an InSb nanowire in proximity with a NbTiN superconductor [see \fref{Kouwenhoven_dIdV}(a)] and the differential conductance  is measured as a function of the magnetic field. As proposed  in Ref.~[\onlinecite{Flensberg:2010}]  electrons from the normal metal scatter against the zero energy MF thus revealing a zero bias anomaly in the differential conductance.  In \fref{Kouwenhoven_dIdV}(b) we can see that as soon as the magnetic field is large enough to satisfy the topological criterion, a zero bias peak appears that is interpreted as the resonant scattering of electrons through the zero energy Majorana state. 
Experiments on these kind of structures are very challenging: on the one hand in order to satisfy the topological criterion, a sizable magnetic field is required, on the other hand, the same magnetic field can disrupt the very fragile superconductivity induced in the semiconducting nanowire by the proximity effect~\cite{Potter:2011}. Everything has to be finely tuned and the temperature has to be quite low, usually below 100mK. 
That is why  many other proposals involving high critical temperature superconductors~\cite{Mezzacapo:2012,Lucignano:2013} or completely different platforms~\cite{Nadj-Perge:2014,Hart:2014} have been investigated, too.

\section{Conclusions and outlook}\label{conclusions}
In this review, we have inspected some of the most interesting theoretical and experimental results on spin dependent quantum transport in mesoscopic systems in the presence of RSOI.
During the last decades a significant number of theories and experiments have appeared. The main goal of this review is to give an historical overview of the field so to address the reader directly to a more specialistic literature. 
Rashba spin-orbit interaction turned out to be  an invaluable tool to have access and to manipulate the electron spin degree of freedom, without breaking time reversal symmetry.  From the applicative point of view, RSOI has been intensively studied to generate spin currents in semiconducting electron 
systems. Several mechanisms and devices for producing pure spin currents have been put forward. Among them, here we have  focused our attention on quantum spin ratchet and spin pumping.

The physics associated to the RSOI can have very applicative oriented aspects but can also be of stimulus for addressing more fundamental issues of quantum mechanics, \emph{e.g.}, quantum interference effects. These have been shown to be strongly affected by the RSOI, with the novelty of the phase factors of non-Abelian nature. We have seen how this additional phase can give rise to  unexpected phenomena, such as the localisation in networks with nontrivial connectivity, or the appearance of anomalous peaks in the Fourier transform of the magneto conductance in quantum rings. 
Recently, new and exciting perspectives open, because RSOI is of fundamental relevance for the physics of topological insulators and of Majorana fermions in hybrid superconducting-semiconducting heterostructures that are currently attracting a lot of interest not only in the condensed matter community.

{Besides charge  and spin transport, SOI may have a relevant role also in heat transport  which is the subject of  \emph{spin caloritronics}~\cite{Bauer:2012}. Here, the combination of magnetic and non magnetic hybrid structure can lead to anomalous behaviour in  the Seebeck and/or the Peltier coefficients~\cite{Gravier:2006,Flipse:2012}. Deviation from standard behavior has been observed also in experiments on  SLG ~\cite{Zuev:2009,Wei:2009}. Furthermore, it is well known that the most efficient media for creating thermoelectric devices are based on materials with a very strong SOI as HgTe, PbTe, Bi,Te, and Bi/Sb alloys,~Ref.~\cite{Delves:1965,Biswas:2012}. However, a complete understanding of the role of RSOI on these properties is still under investigation~\cite{Alomar:2014,Rameshti:2015}.}

We are sure that in the future this field of research will be still thriving. Most probably new electronic and spintronic  devices will come to the market with functionality that are associated to the physics of the RSOI. On the other side, as it happened for the case of the Majorana quasiparticle research, RSOI could pave the way to fundamental research in order to investigate exotic phenomena that could get accessed by \emph{quantum simulators} realised in solid state systems~\cite{FEYNMAN:1982}.

\begin{acknowledgments}
We gratefully acknowledge all our collaborators on the subject of quantum transport in systems with Rashba spin-orbit interaction. In particular F.S. Bergeret, D. Gos\'albez Mart\'inez, V. Marigliano Ramaglia and A. Tagliacozzo that have carefully read the manuscript. D.B. acknowledges financial support from Ikerbasque Foundation of Science and the Alexander von Humboldt Foundation. P.L. acknowledges financial support from  the FIRB project HybridNanoDev RBFR1236VV of Italian ministry of education.
\end{acknowledgments}

\bibliographystyle{apsrev4-1}
\bibliography{reviewBibliography}

\end{document}